\documentclass[preprint,12pt]{elsarticle}

\usepackage{lscape}

\usepackage{graphicx}

\usepackage{amssymb}
\usepackage[mathlines]{lineno}

\begin{document}

\begin{frontmatter}

\title{Regolith grain sizes of Saturn's rings inferred from Cassini-CIRS far-infrared spectra}

\author[jpl,ucla]{Ryuji Morishima\corref{cor1}}
\ead{Ryuji.Morishima@jpl.nasa.gov}

\author[jpl]{Scott G. Edgington}

\author[jpl]{Linda Spilker}

\cortext[cor1]{Corresponding author}

\address[jpl]{Jet Propulsion Laboratory/California Institute of Technology,
Pasadena, CA 91109, USA}

\address[ucla]{University of California, Los Angels, Institute of Geophysics and Planetary Physics, Los Angeles, CA  90095, USA}

\begin{abstract}
We analyze far-infrared (10-650 cm$^{-1}$) emissivity spectra of Saturn's main rings obtained by 
the Cassini Composite Infrared Spectrometer (CIRS). 
In modeling of the spectra, the single scattering albedos of regolith grains are calculated using the Mie theory, diffraction
is removed with the delta-Eddington approximation, and the hemispherical emissivities of macroscopic free-floating ring particles 
are calculated using the Hapke's isotropic scattering model.
Only pure crystalline water ice is considered and  the size distribution of regolith grains is estimated. 
We find that good fits are obtained if the size distribution is broad ranging from 1 $\mu$m to 1-10 cm  with a power law index of $ \sim 3$.
This means that  the largest regolith grains are comparable to the smallest free-floating particles in size and that
the power law indices for both free-floating particles and regolith grains are similar to each other.
The apparent relative abundance of small grains increases with decreasing solar phase angle (or increasing mean temperature).
This trend is particularly strong for the C ring and is probably caused by eclipse cooling in Saturn's shadow, 
which relatively suppresses warming up of grains larger than the thermal skin depth ($\sim$ 1 mm) under subsequent solar illumination. 
\end{abstract}

\end{frontmatter}


\section{Introduction}
Saturn's rings consist of a large number of icy particles. 
The range of the particle size in the main rings (the A, B, C rings, and the Cassini division)  
deduced from radio and stellar light occultations is roughly 1 cm to 10 m 
(Marouf et al., 1983; Zebker et al., 1985; French and Nicholson, 2000; Cuzzi et al., 2009).
The composition of ring particles is mostly crystalline water ice and the mass fraction of contaminants
(e.g., Tholins, PAHs, or nanohematite) is ten percent at most (Epstein et al., 1984; Cuzzi et al., 2009)
 and probably less than one percent (Poulet et al., 2003). 
A favorable origin of Saturn's rings with such a high content of water ice is stripping of the icy mantle of  a Titan-sized  satellite (Canup, 2010).
Subsequent meteoritic pollution for 4.5 Gyr may darken the rings too much if
the ring mass has been kept similar to the present mass (Cuzzi and Estrada, 1998; Elliot and Esposito, 2011), 
but initially very massive rings suggested from the Canup's model have not yet been taken into account in pollution models.

Individual ring particles are likely to be covered by regolith grains
\footnote{Throughout the paper, we use "particles" for free-floating particles and "grains" for regolith grains.},
probably formed by meteoritic bombardment (Elliot and Esposito, 2011).  
The regolith grain size is estimated from ring spectra at different wavelengths from ultraviolet to submillimeter ranges,  
but the estimated size is puzzling as it 
increases with wavelength.
For far-ultraviolet wavelengths, Bradley et al. (2010)  estimate the mean photon path length, which is probably the order of 
the grain size,  as ~2-5 $\mu m$. For near-infrared wavelengths,  
the estimated grain size varies from author to author
(Nicholson et al., 2008; Cuzzi et al., 2009\footnote{A work by R. Clark in Cuzzi et al. (2009)}; Filacchione et al., 2012)
ranging from 5 to 100 $\mu m$, but all these works suggest a relatively smaller size in the C ring than the A and B rings.
These works for near-infrared spectra assume that the ring composition is pure water ice, and  
the size difference could be due to different water bands used or different radiative transfer models. 
Poulet et al. (2003)  study the ring composition using both visible and near infrared spectra. 
Their model calls for a wide spread in grain sizes: 10-1000  $\mu m$ for the A and B rings and 30-7500 $\mu m$ for the C ring.
The large difference in the grain size of the C ring from other works indicates that there is some degeneracy between the ring composition and grain size.
For far-infrared wavelengths,
Spilker et al. (2005) find that  the B ring spectrum is well fitted by a model spectrum 
with the grain sizes ranging from 8 $\mu m$ to 10 m and the power-law index of 3.4 
(the upper limit of 10 m means that the entire ring is considered as regolith layers in their study).
 
Since Saturn orbit insertion of the Cassini spacecraft in July 2004, the Cassini Composite Infrared Spectrometer (CIRS) 
has obtained millions of spectra (7 $\mu$m - 1 mm) of Saturn's rings (Flasar et al., 2005; Spilker et al., 2005, 2006; 
Altobelli et al., 2007, 2008, 2009; Leyrat et al., 2008b; Flandes et al., 2010, Morishima et al., 2011). 
Except for Spilker et al. (2005),  all above works discuss ring temperatures. 
The signal-to-noise levels of individual spectra are usually 
good enough to derive temperatures using Planck fits 
whereas averaging over many spectra are necessary for spectroscopy to examine grain sizes and 
possible contaminants. 
In Spilker et al. (2005), only the spectral data in the very early phase of the mission are analyzed.
In the present study, 
using the many spectra we have obtained so far, radial variation and  dependence on observational geometries are examined in detail. 

In Section~2, we explain how to derive emissivity spectra from observed radiances.
In Section~3, selection and averaging of data are discussed. 
In Section~4, the modeling of emissivity spectra is described.
In Section~5, the estimated grain size distributions for different rings are shown. 
Interpretations of the results and comparison with previous works are discussed in Section~6.
The summary of the present work is given in Section~7. 
 
\section{Derivation of emissivity spectra from observed radiances}
The quantity directly obtained by the spectrometer is the radiance $I(\nu)$
as a function of wavenumber, $\nu$. If the emissivity $\epsilon(\nu)$ is assumed not to vary over a footprint, 
the form of $I(\nu)$ for ring thermal emission is given as 
\begin{equation}
I(\nu) = \epsilon(\nu)\beta_{\rm geo}\int f(T) B(T,\nu) dT, \label{eq:radi} 
\end{equation}
where $\beta_{\rm geo} (\le 1)$ is the geometric filling factor of a ring (see Sec.~2.4),
$f(T)$ is the distribution function of temperature $T$ inside a footprint, normalized as  $\int f(T) dT = 1$,
and $B(T,\nu)$ is the Planck function.
Because the temperature inside the footprint is not known and may not be uniform,
it is not straightforward to derive 
$\epsilon(\nu)$  from $I(\nu)$, unlike temperature-controlled laboratory measurements. 
With different assumptions, 
different types of emissivities are obtained. Three different 
emissivities are introduced in the following: $\epsilon_0(\nu)$ derived by
a fit with a single temperature, $\epsilon_1(\nu)$ derived by 
a fit with two temperatures, and  $\epsilon_2(\nu)$ derived by 
a fit with two temperatures and an expected emissivity spectrum.

\subsection{Fits with a single temperature}
If the temperature distribution is represented by a single
effective temperature $T_0$, one may have 
\begin{equation}
\int f(T) B(T,\nu) dT = \beta_{\rm therm}(\nu) B(T_0,\nu), \label{eq:int0}
\end{equation}
where $\beta_{\rm therm}$ is the correction factor due to multiple temperatures inside a footprint.
The expected shape of  $\beta_{\rm therm}$ is discussed in Sec.~2.2.

In the first method to obtain the emissivity is to assume $\beta_{\rm therm}(\nu) = 1$.
Substituting Eq.~(\ref{eq:int0}) into Eq.~(\ref{eq:radi}), we obtain
\begin{equation}
\epsilon_{0}(\nu) = \frac{I(\nu)}{\beta_{\rm geo} B(T_0,\nu)}.\label{eq:e0}
\end{equation} 
The temperature $T_0$ is obtained by minimizing the 
residuals $R_0$
\begin{equation}
R_0 (T_0,  \epsilon_{\rm sca,0}) =  \sum_{\nu} \left(\frac{I(\nu) - \epsilon_{\rm sca,0}\beta_{\rm geo}B(T_0,\nu)}{\sigma(\nu)}\right)^2, \label{eq:res0}
\end{equation}
where  $\epsilon_{\rm sca,0}$ is the (scalar) emissivity averaged over wavelengths and $\sigma(\nu)$ is the instrument noise equivalent spectral radiance (NESR; see Flasar et al. 2004).
The above Planck fits are applied to all CIRS ring spectra to obtain corresponding temperatures. 
In Spilker et al. (2005), Eqs.~(\ref{eq:e0}) and (\ref{eq:res0}) are used and $\epsilon_{0}(\nu)$ is averaged over many spectra.
Altobelli et al. (2008) estimated $\epsilon_{\rm sca,0}$ for the C ring to be $\sim$ 0.9.
It is expected that Eq.~(\ref{eq:res0}) works well if  the temperature variation in the footprint is small ($\beta_{\rm therm}(\nu) \simeq 1$) and $\epsilon(\nu)$ is roughly flat over wavelength of interest.  

\subsection{Fits with two temperatures}
It is unlikely that the temperature inside a single footprint is perfectly uniform. The particle 
temperature may vary radially and vertically. Even on  the surface of a single particle,
large temperature variation may exist as 
the temperature is expected to be highest and lowest around
the sub-solar and anti-sub-solar points, respectively.
Instead of  
deriving a temperature distribution with complicated thermal models which calculate the energy balance of rings,
simply two portions with high and low temperatures ($T_{\rm w1}$ and $T_{\rm c1}$) are considered:
\begin{equation}
\int f(T) B(T,\nu) dT = f_{\rm w1} B(T_{\rm w1},\nu) + (1- f_{\rm w1}) B(T_{\rm c1},\nu), \label{eq:int1}
\end{equation}
where $f_{\rm w1}$ is the fraction of the warm portion.
Fits with two temperatures are also used in analysis of thermal emission of icy satellites (Carvano et al. 2007; Howett et al. 2011). 
Substituting Eq.~(\ref{eq:int1}) into Eq.~(\ref{eq:radi}), the emissivity is given by 
\begin{equation}
\epsilon_{\rm 1}(\nu) =  \frac{I(\nu)}{ \beta_{\rm geo}\left[f_{\rm w1} B(T_{\rm w1},\nu) + (1-f_{\rm w1}) B(T_{\rm c1},\nu)\right]}.\label{eq:e1}
\end{equation} 
As in Eq.~(\ref{eq:res0}), the temperatures and factors are determined by minimizing the residual as
\begin{eqnarray}
&& R_1 (T_{\rm w1}, T_{\rm c1},\epsilon_{\rm sca,1}, f_{\rm w1}) =  \nonumber \\
&& \sum_{\nu} \left(\frac{I(\nu) - \epsilon_{\rm sca,1}\beta_{\rm geo}\left[f_{\rm w1} B(T_{\rm w1},\nu) + (1-f_{\rm w1}) B(T_{\rm c1},\nu)\right]}{\sigma(\nu)}\right)^2, \label{eq:res1}
\end{eqnarray}
where  $\epsilon_{\rm sca, 1}$ is the scalar emissivity. 
For the convenience of later use, we also define 
the relative emissivity $\epsilon_{\rm rel,1}(\nu)$ as 
\begin{equation}
\epsilon_{\rm rel,1}(\nu) = \frac{\epsilon_{\rm 1}(\nu)}{\epsilon_{\rm sca, 1}}.\label{eq:er1}
\end{equation} 
The relative emissivity is roughly scaled to be unity, so is suitable for comparison between different spectra.

Equating Eqs.~(\ref{eq:int0}) and (\ref{eq:int1}), the form of $\beta_{\rm therm}$ is given as
\begin{equation}
\beta_{\rm therm}(\nu) = \frac{f_{\rm w1} B(T_{\rm w1},\nu) + 
(1-f_{\rm w1}) B(T_{\rm c1},\nu)}{B(T_0,\nu)}. \label{eq:bth}
\end{equation}
An example of $\beta_{\rm therm}$ for various values of $f_{\rm w1}$ 
is shown in Fig.~1.  Near the Planck peak ($\sim$ 200 cm$^{-1}$), $\beta_{\rm therm}$ takes the lowest value whereas
$\beta_{\rm therm}$ is much larger than unity at large $\nu$. Indeed most of spectra show large $\epsilon$ at large 
$\nu$ due to the effect of $\beta_{\rm therm}$, if fits are done with a single Planck function.

\subsection{Fits with two temperatures and an expected emissivity}
The method in Sec.~2.2 works only if $\epsilon(\nu)$ is globally flat with possible narrow features 
in the wavenumber range of interest. If broad features exist, correct temperatures may not be retrieved. 
An extreme case is that $\epsilon(\nu)$ has a profile similar to $ 1/\beta_{\rm therm}(\nu)$.
In this case, both  $\epsilon(\nu)$ and $\beta_{\rm therm}(\nu)$ are incorrectly found to be nearly
constant over the range of $\nu$. In fact, we have this problem because
$\epsilon(\nu)$ for water ice is an upward convex function around the Planck peak wavenumber for the ring temperature (Sec.~4). 
In this kind of situation, the temperatures ($T_{\rm w2}$ and $T_{\rm c2}$) and the fraction of the warm portion ($f_{\rm w2}$)
may be retrieved with an expected (modeled) emissivity $\epsilon_{\rm model} (n_1,n_2,..., \nu)$,
where $n_1$, $n_2$, ... are parameters in models, as
\begin{eqnarray}
R_2(T_{\rm w2}, T_{\rm c2},f_{\rm w2},n_1,n_2,...) = \sum_{\nu} \left(\frac{I(\nu)  - I_{\rm model}(T_{\rm w2}, T_{\rm c2},f_{\rm w2},n_1,n_2,...,\nu)}{\sigma(\nu)}\right)^2, \label{eq:res2}
\end{eqnarray}
where
\begin{eqnarray}
&& I_{\rm model}(T_{\rm w2}, T_{\rm c2},f_{\rm w2},n_1,n_2,...,\nu) = \nonumber \\
&&  \epsilon_{\rm model}(n_1,n_2,...,\nu)\beta_{\rm geo} \left[f_{\rm w2} B(T_{\rm w2},\nu) + (1- f_{\rm w2}) B(T_{\rm c2},\nu)\right].
\end{eqnarray} 
The retrieved emissivity is given as
\begin{equation}
\epsilon_{\rm 2}(\nu) =  \frac{I(\nu)}{\beta_{\rm geo}\left[f_{\rm w2} B(T_{\rm w2},\nu) + (1-f_{\rm w2}) B(T_{\rm c2},\nu)\right]}.\label{eq:e2}
\end{equation} 
It is expected that $\epsilon_{\rm model}(\nu)$ with best-fit parameters is close to $\epsilon_{\rm 2}(\nu)$.

In this study, we estimate best-fit values of model parameters using Eq.~(\ref{eq:res2}), rather than comparing $\epsilon_1(\nu)$ with $\epsilon_{\rm model}(\nu)$,
because only poor fits are obtained for the latter case.

\subsection{Geometric filling factor}
The geometric filling factor $\beta_{\rm geo}$ is given as 
\begin{equation}
\beta_{\rm geo}(B_{\rm o}) = 1-\exp\left(-\frac{\tau(B_{\rm o})}{\sin{|B_{\rm o}|}}\right), \label{eq:bgeo}
\end{equation} 
where  $B_{\rm o}$ is the elevation angle of the observer and $\tau(B_{\rm o})$ is the normal optical depth of the ring.
The optical depth $\tau$ is known to depend on $B_{\rm o}$ and also on longitude of the observer around the footprint relative to 
the ring longitude because of wakes (Colwell et al., 2006, 2007, 2010; Hedman et al., 2007).
In the present study, we only consider the $B_{\rm o}$-dependence and $\tau(B_{\rm o})$ is obtained by interpolations using  
five $\tau$ profiles with different values of $B_{\rm o}$ (15.3, 21.1, 28.7, 54.0, and 66.7 degs).
These data are available in Planetary Data System (Colwell et al., 2010).
The $\tau$ profile for $B_{\rm o} = 28.7$ degs is obtained by Voyager PPS while others are by 
Cassini UVIS.
We interpolate $\tau$ if $\tau (B_{\rm o} =$ 28.7 degs) $< 0.2$ then convert it to $\beta_{\rm geo}$, whereas $\beta_{\rm geo}$ is 
interpolated for larger $\tau$. Usually $\tau$ has little dependence on $B_{\rm o}$ in regions with small $\tau$.
The interpolated profile of $\beta_{\rm geo}$ is smoothed over CIRS footprints (see Morishima et al., 2010).

\section{Selection and averaging of data}
Individual CIRS spectra are usually too noisy for spectroscopy and averaging over many spectra is necessary to reduce 
the noise level. Different types of averaging are considered. In previous works (Spilker et al., 2005; Carvano et al., 2007), 
the emissivities ($\epsilon_{0}(\nu)$ and $\epsilon_{1}(\nu)$) are averaged over after applying Planck fits to individual spectra. 
On the other hand, the radiance is averaged over and Planck fits are applied to the averaged spectra in this study.  
There are advantages and disadvantages in both methods. The latter method is numerically less intense than 
the former method, particularly if the number of parameters used in Planck fits are numerous like the case of $\epsilon_{2}(\nu)$.
The channel-to-channel noise level is smaller for the latter method than the former method, as the noise originally arises in the radiance.  

However, the following conditions are necessary for the latter method to work: 
(1) variation of the emissivity is small enough in the spectra used in an average, (2) correlation between 
 $\beta_{\rm therm} B(T,\nu)$ (temperature and its variation) and $\beta_{\rm geo}(B_{\rm o})$ is small enough, and 
(3) temperature variation including both inside of individual footprints and over different footprints are still 
well approximated  by a sum of two Planck functions. Averaging Eq.~(\ref{eq:radi}) over spectra, these conditions are represented as: 
\begin{eqnarray}
 \langle I(\nu) \rangle &=& \epsilon_1(\nu) \langle  \beta_{\rm therm}(\nu)B(T,\nu) \beta_{\rm geo} \rangle \nonumber \\
 &= & \epsilon_1(\nu) \langle  \beta_{\rm therm}(\nu)B(T,\nu)\rangle \langle \beta_{\rm geo} \rangle \nonumber \\ 
 &= & \epsilon_1(\nu) \left[ f_{\rm w1} B(T_{\rm w1},\nu) + (1-f_{\rm w1}) B(T_{\rm c1},\nu)\right]   \langle \beta_{\rm geo} \rangle.
 \end{eqnarray}
A similar equation is also obtained for $ \epsilon_2(\nu)$.

In order to check whether these conditions are fulfilled,  we test various types of grouping of spectra 
based on saturnocentric radius, observational geometry, and temperature, $T_0$, while keeping a certain number of spectra 
in a parameter bin for a good signal-to-noise level.  
After many trials, we divide spectra into six radial bins (the inner and outer C ring, inner and outer B ring, Cassini division, and A ring) 
and into temperature ($T_0$) bins with the interval of 5 K. These bins are chosen because apparent spectral shapes 
depend on saturnocentric radius and temperature. The temperature dependence probably represents the 
dependence on observational geometries. 
Noise cancellation works better with temperature bins 
rather than using bins of geometric parameters, and also
an unnecessary enhancement of the effect of $\beta_{\rm therm}$  is suppressed 
by averaging spectra with similar temperatures.
We do not  divide spectra by $B_{\rm o}$ because we 
find that there is only small difference between averaged spectra at high and low $|B_{\rm o}|$, provided that other conditions are the same. 
This means that the condition (2) is almost fulfilled.  
The number of spectra in the bins, the mean solar phase angle, and 
the mean absolute solar elevation angles are shown in Fig.~2. The highest temperature is obtained at low phase angle and at high solar elevation,
while the temperature decreases with decreasing solar elevation and increasing phase angles.

The above binning is made for all spectra obtained until the end of 2010 after removing very noisy spectra.
We only analyze spectra in the far-infrared channel (focal plane 1;  10 cm$^{-1} < \nu < $ 650 cm$^{-1}$) with $T_0 \ge $ 70 K, 
as spectra at low temperatures or in the mid-infrared channel (focal planes 3 and 4; 600 cm$^{-1} < \nu < $ 1400cm$^{-1}$) are too noisy even after averaging.
We only show spectral data at the lowest spectral resolution, 15.5 cm$^{-1}$ (the interpolated spectral step is 5 cm$^{-1}$), in this paper.
Spectra with higher resolutions are noisier but show good agreement with the lowest resolution data in most cases.  
 
 Examples of the relative emissivity $\epsilon_{\rm rel,1}(\nu)$ for different temperatures $T_0$ are 
 shown in Figs.~3 and 4 for the inner C ring and the outer B ring, respectively. 
The emissivity spectra for the inner C ring show strong dependence on temperature whereas the inner B ring spectra show 
little dependence on temperature.
 These temperature dependence and independence are also confirmed in higher spectral resolution data.
 The dip seen at  230 cm$^{-1}$ (clearly seen in Fig.~4 but ambiguous in Fig.~3) 
 is a feature caused by intermolecular vibrations (transverse-optical mode) of water ice.  
 Other narrow features are noises specific to the CIRS instrument 
  (Carlson et al. 2009\footnote{De-spiking of interferograms described in Carlson et al. (2009) has not been applied 
  to the 15.5 cm$^{-1}$ data. See also "Interferences on CIRS interferograms and spectra: A user guide" 
 by Nixon et al. at http://pds-rings.seti.org/vol/COCIRS\textunderscore0409/DOCUMENT/cirs\textunderscore
 interferences.pdf.}): the real-time interrupt (RTI) noise  (191, 383, and 574 cm$^{-1}$) and a prominent single frequency 
(sine wave in interferograms) feature of unknown origin that has been observed between 18 cm$^{-1}$ and 210 cm$^{-1}$ 
(seen at 80 cm$^{-1}$ in Fig.~3 and at 150 cm$^{-1}$ in Fig.~4).  
 The noise level is not shown in Figs.~3 and 4, but  the channel-to-channel noise level is expected to be
  $\sim \sigma (\nu)/(\langle \beta_{\rm geo}\rangle B(T_0,\nu)N_{\rm spec}^{1/2})$ 
 and is small enough due to large $N_{\rm spec}$, except at the wave numbers of the CIRS specific noise.   
 The standard deviation of the radiance in spectral averaging is not used, as it is independent of $N_{\rm spec}$ 
 and makes formal error bars artificially large (Clark et al., 2008).
 Most of the spectra  in Figs.~3 and 4 show increase of $\epsilon_{\rm rel,1}(\nu)$ with $\nu$ for $\nu > 400 $ cm$^{-1}$. 
 This is due to the effect of   $\beta_{\rm therm} $ which is not removed appropriately
even with two temperature fits. This artificial effect will be better corrected in $\epsilon_2(\nu)$.
   
\section{Emissivity modeling}
In our emissivity model, the single scattering albedos of regolith grains are calculated using the Mie theory (Mie, 1908; Bohren and Huffman, 1983), 
diffraction is removed with the delta-Eddington approximation (Joseph et al., 1974), 
and the hemispherical emissivities for macroscopic particles 
are calculated using the Hapke's isotropic scattering model (Hapke, 1993).
Cheng et al. (2010) compare  the 
mid-infrared emissivity spectra (8-13 $\mu$m) of snow surfaces obtained by field measurements of Hori et al. (2006)
with those from various types of emissivity models, and find that 
the Mie-Hapke hybrid model with diffraction removal is the best choice. 
The Mie-Hapke hybrid model is also the best choice for emissivity modeling of quartz (Moersch and Christensen, 1995) and 
olivine (Mustard and Hays, 1997), although diffraction removal does not improve fits for quartz (Pitman et al., 2005).

The Mie theory calculates absorption and scattering of a sphere with a given size parameter $X = 2\pi r/ \lambda$
 (where $r$ is the grain radius and $\lambda$ is the wavelength of incident wave) 
and the real and imaginary parts of the refractive indices, $n$ and $k$.
In the present work, we only consider pure crystalline water ice.
To date non-water ices (NH$_3$, CO$_2$, and CH$_4$) have not yet been identified in CIRS far-infrared spectra (Edgington et al., 2008),
and effects of possible contaminants will be examined more in detail in future work.
The refractive indices of crystalline water ice used for the Mie calculations are shown in Fig.~5. 
We use $n$ and $k$ at 136 K from Curtis et al. (2005) for 50 cm$^{-1} < \nu < $ 650cm$^{-1}$
and $k$ at 90 K  from the theoretical fitting function of Mishima et al. (1983) for 10 cm$^{-1} < \nu < $ 30 cm$^{-1}$. 
We assume that $n$ is constant for $\nu < $ 50 cm$^{-1}$, and $\log{k}$ is linearly interpolated between 30 and 50 cm$^{-1}$.
The temperature of Saturn's rings is lower than 136 K, but no good data are available at ring temperatures 
for crystalline water ice. The refractive indices weakly depend on temperature,
particularly, $\nu$ at the peak of $k$ (near 230 cm$^{-1}$) weakly increases with decreasing temperature (Curtis et al., 2005).
In fact, the CIRS emissivity spectra show the peak wavenumber slightly larger than the model prediction (see Fig.~8), 
as we use $n$ and $k$ at the warmer temperature. 
The samples of Curtis et al. (2005) at temperatures lower than 136 K are amorphous and 
amorphous ice shows $\nu$ at the peak of $k$ slightly lower than that for crystalline ice (Fig.~5).
This indicates that Saturn's rings most likely consist of crystalline ice.  
The same conclusion is also derived from the Fresnel peak position in near-infrared spectra (Cuzzi et al., 2009).

Using the extinction efficiency $Q_{\rm ext0}$ and the scattering efficiency $Q_{\rm sca0}$ from the Mie theory, 
the single scattering albedo $\omega_0$ is given as 
\begin{equation}
\omega_0 = \frac{Q_{\rm sca0}}{Q_{\rm ext0}}.
\end{equation}
Figure~6a shows $Q_{\rm ext0}$ and $\omega_0$ for various grain radii.
In the Mie theory, scattered light includes diffracted light. 
In a closely packed medium, however, inter-grain spaces are narrow so that the diffracted light is suppressed (Hapke, 1999).  
Different methods of diffraction removal are proposed (Joseph et al., 1976; Wiscombe, 1977; Mishchenko, 1994; Wald, 1994) 
and we apply the delta-Eddington approximation proposed by Joseph et al. (1976) because of its convenience. In this approximation, 
the diffraction is approximated by a forward-scattering delta function and its contribution is calculated by assuming that 
the second moment of the sum of the delta function and the diffraction-removed phase function
is identical to that for the original phase function. 
The extinction efficiency $Q_{\rm ext}$, the scattering efficiency $Q_{\rm sca}$, and the single scattering albedo, $\omega$, after diffraction removal are given as 
\begin{eqnarray}
Q_{\rm ext}  &=& (1-\omega_0 g_0^2) Q_{\rm ext0},\\
Q_{\rm sca}  &=& (1- g_0^2)\omega_0 Q_{\rm sca0},\\
\omega &=& \frac{Q_{\rm sca}}{Q_{\rm ext}},
\end{eqnarray}
where $g_0$ is the anisotropic parameter calculated by the Mie-theory.
Figure~6b shows $Q_{\rm ext}$ and $\omega$ for same grain sizes used in Fig.~6a.
It is found that $\omega$ is usually much lower than $\omega_0$. 
For $X \gg 1$, $Q_{\rm ext0} \simeq 2$ whereas $Q_{\rm ext} \simeq 1$, as expected.
The Mie parameters are little affected by the delta-Eddington approximation
for small grains $(X < 1)$,  because $g_0$ approaches zero in the Rayleigh scattering regime.
However, differences in $\omega$ due to different diffraction subtraction methods are large for small grains (not shown). 
Fortunately, this has little impact if a broad size distribution including large grains is taken into account, because $Q_{\rm ext}$
for small grains is so small that its contribution to scattering is reduced (see Eq.~(\ref{eq:wa})). 
The exception is near the peak of $k$ at 230 cm$^{-1}$, where $Q_{\rm ext}$ is large even for $X < 1$, and 
there is little difference between $Q_{\rm ext0}$ and $Q_{\rm ext}$, even applying any diffraction subtraction method.

For a size distribution, the single scattering albedo is effectively given as 
\begin{equation}
\omega = \frac{\int \pi r^2 Q_{\rm sca}(r) n(r) dr} {\int \pi r^2 Q_{\rm ext}(r) n(r) dr}. \label{eq:wa}
\end{equation}
We assume the number of grains per unit radius is given by a power-law 
\begin{equation}
 n(r) = n_0\left(\frac{r}{r_0}\right)^{-p}, 
\end{equation}
where $p$ is the index, $n_0$ and $r_0$ are constants.
The black thin lines in Fig.~6a and 6b show $\omega_0$ and  $\omega$ for $p = 3$ with the minimum and maximum cut-off sizes
of $r_{\min} = 1$ $\mu$m and $r_{\max} = 1$ cm. The logarithmic bins with an increment of 2$^{1/4}$ are used for $r$. 

Hapke (1993) derives analytic expressions of various types of reflectances and emittances
using the two stream approximation for isotropic scattering grains.
Using $\omega$, the Bond (spherical) albedo of 
a macroscopic ring particle comprised of regolith grains is given as (Eq.~(10.51b) of Hapke, 1993)
\begin{equation}
A_{\rm BOND} = \frac{1-\gamma}{1+\gamma}\left(1-\frac{1}{3}\frac{\gamma}{1+\gamma}\right), \label{eq:bond}
\end{equation}
with 
\begin{equation}
\gamma = \sqrt{1-\omega}.
\end{equation}
Finally, the hemispherical emissivity $\epsilon_{\rm model}$ of the ring particle 
is given from the Kirchhoff's law as  (Eq.~(13.29) of Hapke, 1993)
\begin{equation}
\epsilon_{\rm model} = 1- A_{\rm BOND}. \label{eq:emodel}
\end{equation}
For spatially resolved planetary or satellite surfaces, the directional emissivity is usually used, whereas we use the hemispherical emissivity 
(the  hemispherically averaged directional emissivity) as we see thermal emission from entire hemispheres of ring particles.
We only considered intra-particle regolith radiative transfer, ignoring inter-particle scattering. Indirect evidence that 
inter-particle scattering is unimportant is obtained from a similarity of the emissivity spectra between the B and A rings
despite the optical depth difference between them (see Sec.~5 for a more detailed discussion).

Bottom panels of Figs.~6a and 6b show $\epsilon_{\rm model}$ without and with diffraction removal. 
 For large grains ($r  >$ 100 $\mu$m), the overall profile of $\epsilon_{\rm model}$ is flat down to $\sim 50$ cm$^{-1}$, 
  and a spectral roll-off is seen for lower $\nu$;
 the exact value of $\nu$ below which a roll-off is seen decreases with increasing $r$.  The location of a roll-off seen in 
 CIRS spectra (Figs.~3 and 4) indicates that large grains indeed exist in regoliths of Saturn's rings.
 The prominent dip of $\epsilon_{\rm model}$ corresponding to the $k$ peak is seen at 230 cm$^{-1}$,
 and $\epsilon_{\rm model}$ is slightly larger for $>$ 230 cm$^{-1}$ than for $<$ 230 cm$^{-1}$ 
 (for example, compare values at 170 cm$^{-1}$ and 280 cm$^{-1}$).  
Intermediate-sized grains (10 $\mu$m $< r  <$ 100 $\mu$m) show a broad hump between 100 cm$^{-1}$ and 300 cm$^{-1}$ due to 
inter-molecular vibrations of water ice. The hump features are also seen in many CIRS spectra.
The emissivity of small grains ($r  <$ 10 $\mu$m) are almost unity for $<$ 300 cm$^{-1}$,
because they behave as Rayleigh absorbers. Their extinction efficiencies are high 
only near 230 cm$^{-1}$ so that they reduce the depth of the dip at  230 cm$^{-1}$ produced by large grains.  

The one issue we find in emissivity fits is that a large amount of small grains is necessary to produce the 
shallow 230 cm$^{-1}$ feature seen in CIRS spectra whereas too large a fraction of small grains makes entire fits worse. 
An alternative way to reduce the depth of the 230 cm$^{-1}$ feature instead of adding too many 
small grains is to introduce intra-grain pores. If the pore size is sufficiently smaller than the wavelength,
the complex permittivity, $K(\phi)$ = $(n^2-k^2) + i(2nk)$,  as a function of the porosity $\phi$ is given from 
the Maxwell-Garnett theory (Maxwell-Garnett, 1904) as 
\begin{equation}
K(\phi) = K(0)\left(1+\frac{3\phi(1-K(0))}{1+2K(0)-\phi(1-K(0))}\right).
\end{equation}
The validity of this equation is confirmed by laboratory measurements of silicon
(Labb\'{e}-Lavigne et al., 1998). 
The refractive indices for $\phi = 0.5$ are shown in Fig.~5.
Figure~6c shows  $Q_{\rm ext}$, $\omega$, and $\epsilon_{\rm model}$ for  $\phi = 0.5$
and with diffraction removal.
It is found that the overall shape of $\epsilon_{\rm model}$ in Fig.~6c is similar to that in Fig.~6b
except the depth of  the 230 cm$^{-1}$ feature is shallower with intra-grain pores. 
Since introducing intra-grain pores has an effect similar to reducing grain-size, 
the estimated grain size effectively increases, but the change in estimated size
is very small as long as  $\phi \le 0.5$.
 
\section{Estimated regolith size}

The best-fit values of model parameters are estimated from Eq.~(\ref{eq:res2}).
In the model fits, we fix the minimum grain size and intra-grain porosity at $r_{\rm min} = 1$ $\mu$m and $\phi = 0.5$
and seek the best-fit values of the maximum size $r_{\rm max}$ and the power-law index $p$.
The reason for choosing $r_{\rm min} = 1$ $\mu$m is that existence of micron-sized grains 
is supported from the grain size of spokes ($r \sim 2$ $\mu$m; D'aversa et al., 2010) and 
from UVIS spectra ($r \sim  2$-5 $\mu$m; Bradley et al., 2010).  We do not examine cases with $r_{\rm max}$ larger than 1 m
because this is a typical size of large free-floating particles (French and Nicholson, 2000). 

The residual $R_2$ is calculated as a function of $r_{\rm max}$ and $p$,
each pair of  $r_{\rm max}$ and $p$ having the optimal temperatures and the fraction of the warm portion. 
We introduce the  normalized residual: 
\begin{equation}
\tilde{R_2}(p,r_{\rm max}) =  \frac{R_2}{R_{\rm 2, min}} - 1, 
\end{equation}
where $R_{\rm 2, min}$ is the minimum 
value of  $R_2$ in the parameter space.  Some examples of $\tilde{R_2}$ on the $r_{\rm max}$ vs. $p$ plane 
are shown in Fig.~7. The error bars for $r_{\rm max}$ and $p$ are estimated from 
the maximum and minimum values of these parameters on the curve of  $\tilde{R_2} = 1$.  
It is found that $p$ can be constrained in some ranges whereas only the lower limit  
of $r_{\rm max}$ (3-10 mm for cases in Fig.~7) is well constrained.

Some examples of the best-fit spectrum, $\epsilon_{\rm model}$, are shown together with $\epsilon_{\rm 2}$ and $\epsilon_{\rm 1}$ in Fig.~8.
Note that $\epsilon_{\rm 2}$ and $\epsilon_{\rm 1}$ are obtained from the same radiance data and the difference comes from 
the fitted temperatures and the fraction of the warm portion (Sections 2.2 and 2.3).  
Good agreements between   $\epsilon_{\rm model}$ and $\epsilon_{\rm 2}$ are obtained for all cases 
at least for $\nu < 300$ cm$^{-1}$ and sometimes over all the wavenumber range.
On the other hand, the absolute value of $\epsilon_{\rm 1}$ is much lower than $\epsilon_{\rm model}$ with 
any grain sizes. In addition to that, $\epsilon_{\rm 1}$ for  $\nu \sim $ 170 cm$^{-1}$ is 
comparable to or higher than $\epsilon_{\rm 1}$ for  $\nu \sim$ 280 cm$^{-1}$ whereas 
the opposite trend is seen for $\epsilon_{\rm model}$. 
The temperature $T_{\rm c2}$ is found to be always lower than $T_{\rm c1}$. 
With decreasing temperature from $T_{\rm c1}$ to $T_{\rm c2}$,
the absolute value of $\epsilon$ increases in all wavenumbers and the overall slope $d\epsilon/d\nu$ increases.
Consequently,  $\epsilon_{\rm 2}$ for $\nu \sim 280$ cm$^{-1}$ can be larger than that for $\nu \sim 170$ cm$^{-1}$ as is the case of pure water ice. 
Although lowering only  the temperature of the cold portion makes $\epsilon$ at large $\nu$ (typically $>$ 400 cm$^{-1}$) larger than unity,
$\epsilon$ in this region is reduced by increasing the temperature and fraction of the warm portion 
(note $f_{\rm w2} > f_{\rm w1} $  or/and $T_{\rm w2} > T_{\rm w1}$ in all cases). 
Some cases shows only one temperature component  ($f_{\rm w1}$ = 0) for the derivation of $\epsilon_{\rm 1}$ 
(e.g., the C ring data in Fig.~8), but
existence of two temperature components, as is the case for $\epsilon_{\rm 2}$,
is also physically reasonable considering a possible 
temperature variation over the surfaces of ring particles. 

Figure~9 shows the summary of estimated values of $p$ and $r_{\rm max}$ for all cases.
It is found that $p \simeq 3$ for most of the cases (Fig.~9a) except that  the C ring shows slightly larger values
and the Cassini division shows slightly smaller values at high temperatures.
Interestingly, the values of $p$ are quite close to those for macroscopic particles (French and Nicholson, 2000).
The maximum size $r_{\rm max}$ is estimated to be 1-100 cm in most of cases.
The lower limit of $r_{\rm max}$ is well constrained, and in almost all cases 
$r_{\rm max}$ is larger than 1 mm (Fig.~9b). On the other hand, the upper limit of $r_{\rm max}$ is 
not well constrained from our data.
However,  $r_{\rm max}$ is expected to be smaller than the size of large particles $\sim$ 1 m at least 
by an order of magnitude and probably even smaller in the following reasons.
First,  the total cross sections of any particles are equal in logarithmic bins for $p = 3$, and 
thus the broad size distribution we obtained means that the surface of a free-floating particle 
is not occupied by the largest regolith grains. 
Second, if inter-grain scattering is less than that expected from the radiative transfer model due 
to large grains comparable to the particle in size, the emissivity spectra may become closer to $1-\omega$ instead of 
$1-A_{\rm Bond}$. This clearly makes our fits worse;  in particular, the depth of the 230 cm$^{-1}$ feature will be too deep.
Most likely,  $r_{\rm max}$ is $\sim$ 1 - 10 cm.
This means that the largest regolith grains are comparable to the smallest 
free-floating particles in size. It is likely that 
grains smaller than 1-10 cm adhesively stick to large particles due to the surface energy (Bodrova et al., 2012),
while larger grains are cores of particles hidden under smaller grains, if they exists.

The A and B rings show quite similar values of $p$ and $r_{\rm max}$.
In addition to that, these rings show very similar depths of the 230 cm$^{-1}$ feature.
If the optical depth of a ring is large enough and inter-particle scattering occurs, 
the emissivity of the ring may be calculated from Eqs.~(\ref{eq:bond}) and (\ref{eq:emodel}) 
assuming that the single scattering albedo is now given as the Bond albedo of ring particles. 
The resulting ring emissivity should 
have a shallower depth of the 230 cm$^{-1}$ feature as compared with that without inter-particle scattering, 
from comparison of the depths for $1-\omega$ and $\epsilon_{\rm model}$
in Fig.~6b or 6c.
Therefore, if regolith sizes are the same and inter-particle scattering is important,
the depth of the 230 cm$^{-1}$ feature is expected to decrease with increasing optical depth; 
such a correlation is not seen in spectra ($\tau \sim 2 $ for the outer B ring while $\tau \sim 0.5$ for the A ring). 
Therefore,  inter-particle scattering of far-infrared light is likely to be negligible.
Little inter-particle scattering in near-infrared light is also suggested at low phase angles (Cuzzi et al., 2009).

The observed spectra for the C ring at low temperatures or the Cassini division 
do not clearly show the 230 cm$^{-1}$ feature in contrast to the modeled ones.
Higher spectral resolution data (not shown) seem to show that 
the feature actually exists both for the C ring and the Cassini division but it is narrower and shallower 
than that for the A and B rings (see also Figs.~3 and 4).
What causes the feature less evident is unclear.
It may be due to low signal-to-noise ratios for the optically thin rings, to intra-grain porosities higher than we assume (0.5), or to effects of contaminants including 
amorphous water ice.
Since higher spectral resolution data are noisy, more careful analysis is necessary.

\section{Discussion}
\subsection{Dependence of size distribution on temperature}
What causes the dependence of size distribution on temperature seen in Fig.~9?
This is probably caused by effects of observational geometries in conjunction with 
non-uniform temperature distribution of ring particles over surfaces and depth.
If the temperature of a ring particle is completely uniform, it is expected that 
the emissivity spectrum of the particle is independent of observational geometry.
However, in actual rings, the strongest thermal emission
comes from near the subsolar points (seen at low solar phase angles) of slowly rotating particles at high solar elevation angles.
In addition to that, due to  eclipse cooling in Saturn's shadow, the strongest emission comes from only 
a thin surface layer with a thickness given by the thermal skin depth $\ell$.
If the thermal inertia is $\sim$10 Jm$^{-2}$K$^{-1}$s$^{-1/2}$ as estimated by Ferrari et al. (2005) and Morishima et al. (2011),
$\ell \sim$ 1 mm (e.g., Morishima et al., 2009). This is smaller than the size of the largest regolith grains estimated in Sec.~5. 
Therefore, the largest grains are only partially heated or their temperatures are lower than small grains near the subsolar point.
As a result, thermal emission from the largest grains is suppressed relative to that from small grains at high temperatures. 
At higher solar phase angles and at lower solar elevation angles, variation of the particle temperature with depth
is much smaller so that large grains can have temperatures similar to those for small grains.
The dependence of size distribution on temperature is seen most strongly in the C ring, probably because
the periodic temperature variation due to eclipse cooling is particularly large for the C ring 
as it is close to Saturn (long duration in the shadow) and has low photometric albedo (high temperature outside the shadow).

To reinforce the above discussion,  let us estimate the depth from which thermal emission comes.
The electric skin depth may be simply given by a depth where the optical depth of regolith layers is unity, unless grains are highly reflecting.
The regolith optical depth as a function of depth is given as   
\begin{equation}
\tau(z) = z\int^{r_{\rm max}}_{r_{\rm min}} \pi r^2 Q_{\rm ext}(r,\nu) n(r) dr,
\end{equation}
where $n$ is the particle number density per unit volume and per unit size.
The number density $n$ is defined by the following equation:
\begin{equation}
1-\phi_{\rm inter} = \int^{r_{\rm max}}_{r_{\rm min}} \frac{4}{3} \pi r^3 n(r) dr, 
\end{equation}
where $\phi_{\rm inter}$ is the inter-grain porosity.
Using these equations, the electric skin depth is given as 
\begin{equation}
z(\tau = 1) = \frac{3 r_{\rm eff}}{4 Q_{\rm ext,eff}(1-\phi_{\rm inter})}, \label{eq:eskin}
\end{equation}
where the effective grain radius $r_{\rm eff}$ and the effective extinction efficiency $Q_{\rm ext,eff}$ are given by 
\begin{equation}
r_{\rm eff} =  \frac{\int^{r_{\rm max}}_{r_{\rm min}} \pi r^3 n(r) dr}{\int^{r_{\rm max}}_{r_{\rm min}} \pi r^2 n(r) dr}, \label{eq:reff}
\end{equation}
and
\begin{equation}
Q_{\rm ext,eff}(\nu) =  \frac{\int^{r_{\rm max}}_{r_{\rm min}} \pi r^2 Q_{\rm ext}(r,\nu) n(r) dr}{\int^{r_{\rm max}}_{r_{\rm min}} \pi r^2 n(r) dr}.
\end{equation}
Since $Q_{\rm ext,eff}(\nu)$ is $\sim$ 0.5-1.0,  $z(\tau = 1)$ is larger than $r_{\rm eff}$ by a factor of 2-3, unless 
$\phi_{\rm inter}$ is close to unity.
Figure~10a shows $r_{\rm eff}$ estimated from $p$ and $r_{\rm max}$ from Fig.~9.
It is found that $r_{\rm eff}$ is $\sim $ 1 mm  ($\sim \ell$) at  the highest temperatures and is one or two orders of magnitude larger for 
lower temperatures. 
This indicates that  if the temperature is uniform as is likely to be the case at low temperatures, 
thermal emission come from 1-10 cm in depth. However, 
thermal emission at the highest temperatures comes from a surface layer of 1 mm because 
only the thin layer is heated to the the highest temperature and emission from deeper and colder layers 
contribute much less to the total emission from the surface.

\subsection{Comparison with grain sizes estimated in previous works}
It is not straightforward to compare the size distribution estimated from the present work and 
those from other works, because other works usually consider single sizes or a few discrete sizes.
If spectral behavior is monotonically dependent on grain size, 
a spectrum for a broad size distribution may be similar to the one for the median grain radius $r_{\rm med}$,
which is defined so that the total cross section of particles with $r>r_{\rm med}$ is equal 
to that with $r < r_{\rm med}$:
\begin{equation}
\int^{r_{\rm max}}_{r_{\rm med}} \pi r^2 n(r) dr = \frac{1}{2} \int^{r_{\rm max}}_{r_{\rm min}} \pi r^2 n(r) dr. \label{eq:rmed}
\end{equation}
Figure~10b shows $r_{\rm med}$. For the C ring,  $r_{\rm med}$ strongly depends on temperature,
but as we discussed in Sec.~6.1, the relative fraction of small grains are likely to be overestimated at high temperatures.
Therefore, $r_{\rm med}$ is typically a few hundred microns or larger for all rings. 

The mean photon path length (2-5 $\mu m$) estimated from far-ultraviolet spectra (Bradley et al., 2010)
is clearly much smaller than $r_{\rm med}$. The path length probably represents the size of cracks or 
flaws inside large grains or represents the smallest grain size, as suggested by Bradley et al. (2010). 
The typical value of $r_{\rm med}$ seems to be also larger than the grain sizes ($\le$ 100 $\mu m$)
estimated from the water band depths in near-infrared wavelengths (Nicholson et al., 2008;  Cuzzi et al., 2009; Filacchione et al., 2012). 
It is cautioned that this comparison is not very precise, as some of the water band depths do not depend on 
particle size monotonically (see Fig.~15.27 of Cuzzi et al. (2009)). Nevertheless, 
some other bands show rather monotonic dependences on size (up to $\sim$ 1-10  cm) and the estimated sizes 
from these bands are still smaller than 100 $\mu m$. This size is probably underestimated as discussed in the following.
 
The grain sizes of snow on Earth are also estimated from reflectances in different bands in visible 
and near-infrared wavelengths and the estimated sizes are compared with the sizes from in-situ measurements.
Some works (Fily et al., 1997; Aoki et al., 2007; Kuchiki et al., 2009) find that the grain sizes 
estimated from the 1.6 $\mu$m band, which is in one of the water absorption bands, are underestimated. 
Kuchiki et al. (2009) find that the grain size is underestimated when the actual grain size is large ($>$ 1 mm) and 
point out that it is probably caused by microstructures on large grains' surfaces. 
It is plausible that regolith grains in ring particles are also not spheres but have surface microstructures. 
If it is the case, it may not be very appropriate to estimate the sizes of large grains from the water band depths,
whereas spectral roll-offs at long wavelengths seen in CIRS spectra are likely to be less affected by microstructures. 
The large C ring grain sizes (30-7500 $\mu m$, which gives $r_{\rm med} \sim$ 1 mm)
estimated from spectra in the entire visible and near-infrared wavelength range (Poulet et al., 2003) look consistent with our results.

\section{Summary}
In this study, the far-infrared emissivity spectra of Saturn's main rings obtained by Cassini CIRS are analyzed. 
Ring spectra are divided into six radial bins and temperature bins with the width of 5 K, and the radiances 
are averaged over bins to reduce the noise levels.
The temperatures and size distribution of regolith grains are simultaneously retrieved 
by the chi-square fits, provided that ring particles are composed of pure water ice.
In modeling,  the Mie-Hapke hybrid model is used with diffraction removal.
Good fits are obtained if size distribution is broad ranging from 
1 $\mu m$ to 1-10 cm  with the power law index of $\sim 3$.
This means that  the largest regolith grains are comparable to the smallest free-floating particles in size and that
the power law indices for both free-floating particles and regolith grains are similar to each other.

The apparent relative abundance of small grains increases with decreasing phase angle (or increasing mean temperature).
This trend is particularly strong for the C ring and is probably caused by eclipse cooling in Saturn's shadow, 
which relatively suppresses warming up of grains larger than the thermal skin depth ($\sim$ 1 mm)  under subsequent solar illumination. 
The A and B ring spectra are found to be very similar to each other despite the difference in optical depth, 
and this probably indicates that there is little scattering of far-infrared light between free-floating macroscopic particles.

There are a lot of things to do in future work. 
Effects of possible contaminants need to be investigated.  Since the imaginary part of the refractive indices for water ice is high in 
the broad range of far-infrared wavelengths,  effects of contaminants are probably small if they are mixed as small inclusions ($r \ll \lambda$) 
inside grains. Nevertheless, some possible contaminants, such as SiO$_2$, Fe$_2$O$_3$, NH$_3$, and NaCl, 
have strong absorption peaks in the CIRS wavelength region,  and the upper limits of their abundances 
may be well constrained even if they are not identified. 
The effects of contaminants are probably more important if they are intimately mixed ($r > \lambda$).  One of plausibly intimately-mixed 
contaminants is amorphous carbon (Poulet et al., 2003).  
It is also necessary to reduce CIRS specific noises with better
calibration. Use of high spectral resolution data may also help improve model fits. 
The accurate determination of refractive indices of crystalline water ice 
at low ring temperatures ($<$ 100 K) are also needed.

\section*{Acknowledgements}
We are grateful to Bruce Hapke and an anonymous reviewer for their thorough reviews.
We thank Jeffrey Cuzzi, Matthew Hedman, John Spenser, Deau Estelle,
and Donald Jennings for fruitful discussions and helpful comments on this work, 
and Stu Pilorz, Shawn Brooks, and Mark Showalter for designing of CIRS observations.
This research was carried out at the Jet Propulsion Laboratory, California Institute of Technology, 
under contract with NASA. Government sponsorship acknowledged. 
This work was also supported in part by NASA's Cassini Data Analysis Program.

\section*{REFERENCES}

\begin{description}

\item
Aoki, T. et al., 2007.
ADEOS-II/GLI snow/ice products - Part II: Validation results using GLI and MODIS data.
Remote Sens. Environ. 111, 274--290

\item 
Altobelli, N., Spilker, L.J., Pilorz, S., Brooks, S., Edgington, S., Wallis, B., Flasar, M., 2007.
C ring fine structures revealed in the thermal infrared.
Icarus 191, 691--701. 

\item 
Altobelli, N., Spilker, L.J., Leyrat, C., Pilorz, S.,  2008.
Thermal observations of Saturn's main rings by Cassini CIRS: 
Phase, emission and solar elevation dependence.
Planet. Space. Sci. 56, 134--146.

\item 
Altobelli, N., Spilker, L.J., Pilorz, S., Leyrat, C., Edgington, S., Wallis, B., Flandes, A., 2009.
Thermal phase curves observed in Saturn's main rings by Cassini-CIRS: Detection of an opposition effect?
Geophys. Res. Lett. 36, L10105.

\item
Bodrova, A., Schmidt, J., Spahn, F., Brilliantov, N., 2012.
Adhesion and collisional release of particles in dense planetary rings.
Icarus 218, 60--68.

\item
Bohren, C.F., Huffman, D.R., 1983.
Absorption and scattering of light by small particles.
Wiley, New York. 

\item
Bradley, E.T. Colwell, J.E., Esposito, L.W., Cuzzi, J.N., Tollerud, H., Chambers, L., 2010.
Far ultraviolet spectral properties of Saturn's rings from Cassini UVIS. 
Icarus 206, 458--466.

\item
Canup, R., 2010.
Origin of Saturn's rings and inner moons by mass removal from a lost Titan-sized satellite.
Nature 468, 943--946.

\item
Carlson, R.C., Guandique, E.A., Jennings, D.E., Pilorz, S.H., Kunde, V.G., 2009.
Characterization and suppression of electrical interference - spikes, periodic waves, and ripples - 
from Cassini Composite Infrared Spectrometer (CIRS) Spectra.
FTS for Astronomy and Astrophysics, FtuA5.

\item
Carvano, J.M., Migliorini, A., Barucci, A., Segura, M., the CIRS team, 2007.
Constraining the surface properties of Saturn's icy moons, using Cassini/CIRS emissivity spectra.
Icarus 187, 574--583.

\item
Cheng, J., Liang, S., Weng, F., Wang, J., Li, X., 2010.
Comparison of radiative transfer models for simulating snow surface
thermal infrared emissivity.
IEEE Journal of Selected Topics in Applied Earth Observations and Remote Sensing, 323--336.

\item
Clark, R.N. et al., 2008.
Compositional mapping of SaturnÕs satellite Dione with Cassini VIMS and implications of dark material in the Saturn system.
Icarus 193, 372--386.

\item 
Colwell, J.E., Esposito, L.W., Srem\v{c}evi\'{c}, M., 2006.
Self-gravity wakes in Saturn's A ring measured by stellar 
occultations from Cassini.
Geophys. Res. Lett. 33, L07201.

\item 
Colwell, J.E., Esposito, L.W., Srem\v{c}evi\'{c}, M., Stewart, G.R., McClintock, W.E.,  2007.
Self-gravity wakes and radial structure of Saturn's B ring.
Icarus, 190, 127--144.

\item
Colwell, J.E., Esposito, L.W., Jerousek, R.J., Srem\v{c}evi\'{c}, M., Pettis, D., Bradley, E.T., 2010.
Cassini UVIS stellar occultation observations of Saturn's rings.
Astron. J. 140, 1569--1578.

\item
Curtis, D.B., Rajaram, B., Toon, O.B., Tolbert, M.A., 2005.
Measurement of the temperature-dependent optical constants of water ice in the 15-200 $\mu$m range.
Appl. Opt. 44, 4102--4118.

\item
Cuzzi, J. N.,  Estrada, P.R., 1998.
Compositional evolution of Saturn's rings due to meteoroid bombardment.
Icarus 132,  1--35.

\item
Cuzzi, J.N., Clark, K., Filacchione, G., French, R., Jhonson, R., Marouf, E., Spilker, L., 2009.
Ring particle composition and size distribution.
In: Dougherty, M.K., Esposito, L.W.,  Krimigis, S.M. (Eds.), Saturn from Cassini-Huygens.
Springer, Berlin, pp. 459--509.

\item 
D'Aversa, E., Bellucci, G., Nicholson, P.D., Hedman, M.M., Brown, R.H., 
Showalter, M.R., Altieri, F., Carrozzo, F.G., Filacchione, G., Tosi, F., 2010. 
The spectrum of a Saturn ring spoke from Cassini/VIMS.
Geophys. Res. L. 37, L01203.

\item
Elliott, J.P., Esposito, L.W., 2011.
Regolith depth growth on an icy body orbiting Saturn and evolution 
of bidirectional reflectance due to surface composition changes.
Icarus 212, 268--274.

\item
Edgington, S.G., Spilker, L.J., Leyrat, C., Nugent, C.R., Jennings, D.E.,  Pilorz, S.H., Pearl, J.C., and the CIRS team, 2008.
Emissivity in the thermal IR: Composition and polarization in Saturn's rings with Cassini/CIRS.
Bulletin of the American Astronomical Society 40, 443

\item
Epstein, E.E., Jamssem, M.A., Cuzzi, J.N., 1984.
Saturn's rings: 3-mm low inclination observations and derived properties.

\item   
Ferrari, C., Galdemard, P., Lagage, P.O., Pantin, E., Quoirin, C., 2005.
Imaging Saturn's rings with CAMIRAS: thermal inertia of B and C rings.
Astron. Astrophys. 441, 379--389.

\item
Filacchione, G., et al., 2012.
Saturn's icy satellites and rings investigated by Cassini-VIMS III: Radial compositional variability.
Icarus 220, 1064--1096.

\item
Fily, M., Bourdelles, B., Dedieu, J.P., Sergent, C., 1997.
Comparison of in situ and Landsat Thematic Mapper derived snow 
grain characteristics in the Alps.
Remote Sens. Environ. 59, 452--460.

\item
Flandes, A., Spilker, L.J., Morishima, R., Pilorz, S., Leyrat, C., Altobelli, N., Brooks, S.M., Edgington, S.G., 2010.
Brightness of Saturn's rings with decreasing solar elevation.
Planet. Space Sci. 58, 1758--1765. 

\item 
Flasar, F.M. et al.,  2004.
Exploring the Saturn system in the thermal infrared: The composite infrared spectrometer.
Space Sci. Rev. 115, 169--297.

\item 
Flasar, F.M. et al., 2005.
Temperatures, winds, and composition in the Saturnian system.
Science 307, 1247--1251.

\item 
French, R.G., Nicholson, P.D., 2000.
Saturn's rings II.
Particle sizes inferred from stellar occultation data.
Icarus 145, 502--523.

\item 
Hapke, B., 1993.
Theory of reflectance and emittance spectroscopy.
Cambridge Univ. Press, New York.

\item
Hapke, B., 1999.
Scattering and diffraction of light by particles in planetary regoliths.
 J. Quant. Spectrosc. Radiat. Transfer 61,  565--581. 
  
\item 
Hedman, M.M., Nicholson, P.D., Salo, H., Wallis, B.D., Buratti, B.J.,
Baines, K.H., Brown, R.H., Clark, R.N.,  2007.
Self-gravity wake  structures in Saturn's A ring revealed by Cassini VIMS.
 Astron. J. 133, 2624--2629.
 
 \item
Hori, M. et al., 2006.
In-situ measured spectral directional emissivity of snow and ice in the 8--14 $\mu$m atmospheric window.
Remote Sens. Environ. 100, 486--502.
  
\item
Howett, C.J.A., Spencer, J.R., Pearl, J., Segura, M., 2011.
High heat flow from Enceladus' south polar region measured using
10-600 cm$^{-1}$ Cassini/CIRS data.
 J. Geo. Res. 116,  E03003.
 
 \item
 Joseph, J.H., Wiscombe, W.J., Weinman, J.A., 1976.
 The delta-Eddington approximation for radiative flux transfer.
 J. Atom. Sci. 33, 2452--2459.
 
 \item
 Kuchiki, K., Aoki, T., Tanikawa, T., Kodama, Y., 2009.
 Retrieval of snow physical parameters using a ground-based spectral radiometer.
 Appl. Opt. 48, 5567--5582.
 
\item
Labb\'{e}-Lavigne, S., Barret, S., Garet, F., Duvillaret, L., Coutaz, J.-L., 1998.
Far-infrared dielectric constant of porous silicon layers measured by terahertz time-domain spectroscopy.
Appl. Phys. 83, 8007--8010.

\item
Leyrat, C., Spilker, L.J., Altobelli, N., Pilorz, S., Ferrari, C., 2008.
Infrared observations of Saturn's rings by Cassini CIRS: Phase angle and local time dependence.
Planet. Space. Sci. 56, 117--133.

\item
Marouf, E.A., Tyler, G.L., Zebker, H.A., Simpson, R.A., 
Eshleman, V.R., 1983. 
Particle size distributions in Saturn's rings from Voyager 1 radio occultation.
Icarus 54, 189--211.

\item
Maxwell-Garnett, J., 1904.
Colours in metal glasses and in metallic films.
Phil. Trans. Roy. Soc. A203, 385--420.

\item
Mie, G., 1908.
Contributions to the optics of turbid media, particularly colloidal metal suspensions (in German).
Ann. Phys. 25, 377--445.

\item
Mishchnko, M.I., 1994.
Asymmetry parameters of the phase function for densely packed scattering grains.
J. Quant. Spectrosc. Radiat. Transfer 52, 95--110.

\item
Mishima, O., Klug, O.O., Whalley, E., 1983.
The far-infrared spectrum of ice Ih in the range 8-25 cm$^{-1}$:
Sound waves and difference bands, with application to Saturn's rings.
J. Chem. Phys. 78, 6399--6404.

\item
Moersch, J.E., Christensen, P.R., 1995.
Thermal emission from particulate surfaces:
A comparison of scattering models with measured spectra.
J. Geo. Phys. 100, 7465--7477.

\item
Morishima, R., Salo, H., Ohtsuki, K., 2009.
A multilayer model for thermal infrared emission of Saturn's rings:
Basic formulation and implications for Earth-based observations.  
Icarus 201, 634--654.

\item
Morishima, R., Spilker, L., Salo, H., Ohtsuki, K., Altobelli, N., Pilorz, S., 2010.
A multilayer model for thermal infrared emission of Saturn's rings II:
Albedo, spins, and vertical mixing of ring particles inferred from Cassini CIRS.  
Icarus 210, 330-345. 

\item  
Morishima, R., Spilker, L., Ohtsuki, K.,  2011.
A multilayer model for thermal infrared emission of Saturn's rings III:
Thermal inertia inferred from Cassini CIRS.
Icarus 215, 107--127. 

\item
Mustard, J.F., Hays, J.E., 1997.
Effects of hyper fine particles on reflectance spectra from 0.3 to 25 $\mu$m.
Icarus 125, 145--163.

\item 
Nicholson, P.D. et al.,  2008.
A close look at Saturn's rings with Cassini VIMS.
Icarus 193, 182--212.

\item
Pitman, K.M., Wolff, M.J., Clayton, G.C., 2005.
Application of modern radiative transfer tools to model laboratory quartz emissivity.
J. Geo. Phys. 110, E08003.

\item
Poulet, F., Cruikshank, D.P., Cuzzi, J.N., Roush, T.L., French, R.G., 2003.
Composition of Saturn's rings A, B, and C from high resolution
near-infrared spectroscopic observations.
Astron. Astrophys. 412, 305--316.

\item
Spilker, L.J., Pilorz, S.H., Edgington, S.G., Wallis, B.D.,
Brooks, S.M., Pearl, J.C., Flasar, F.M., 2005.
Cassini CIRS observations of a roll-off in Saturn ring spectra at
submillimeter wavelengths.  
Earth, Moon, and Planets 96, 149--163. 

\item
Spilker, L.J. et al., 2006.
Cassini thermal observations of Saturn's main rings: 
Implications for particle rotation and vertical mixing.
Planet. Space Sci. 54, 1167--1176.

\item
Wald, A.E., 1994.
Modeling thermal infrared (2-14 $\mu$m) reflectance spectra of frost and snow.
J. Geo. Phys. 99, 24241--24250.

\item
Wiscombe, W.J., 1977.
The delta-M method: rapid yet accurate radiative flux calculations for strong asymmetric phase functions.
J. Atom. Sci. 34, 1408--1422.

\item
Zebker, H.A., Marouf, E.A., Tyler, G.L., 1985.
Saturn's rings - Particle size distributions for thin layer model.
Icarus 64, 531--548.

\end{description}

\clearpage

\section*{Figure captions}

\begin{description}

\item 
Fig.~1.  Examples of $\beta_{\rm therm}$ as a function of $\nu$, given by Eq.~(\ref{eq:bth}). 
Here we adopt $T_{\rm w1}$ = 105 K and  $T_{\rm c1}$ = 80 K, and $T_0$ is obtained from a Planck fit (Eq.~(\ref{eq:res0})). 
Different colors represent different values of the fraction of the warm portion, $f_{\rm w1}$.

\item
Fig.~2. Number of spectra, $N_{\rm spec}$, mean solar phase angle, $\alpha$, and mean absolute solar elevation angle $|B'|$
of spectrum divided in six radial bins and temperature ($T_0$) bins of a 5 K interval.

\item
Fig.~3. The relative emissivity $\epsilon_{\rm rel,1}$ for the inner C ring given by Eq.~(\ref{eq:er1}). 
Different colors represent different temperature bins: 70-75 K(purple), 75-80 K(blue), 80-85(light blue),
85-90 K(brown), 90-95 K(orange), and 95-100 K (red). The upper panel shows the entire wavenumber 
range of the CIRS far-infrared channel, whereas the bottom panel shows the short wavenumber roll-off region. Note the difference in
the vertical scales in these panels.

\item
Fig.~4. The same with Fig.~3, but for the case of the outer B ring.

\item
Fig.~5. 
The real ($n$) and imaginary ($k$) parts of refractive indices of water ice. 
The black solid lines represent $n$ and $k$ for crystalline water ice at 136 K for $\nu > $ 50 cm $^{-1}$ 
(Curtis et al., 2005) and $k$ at 90 K  for 10 cm$^{-1}$ $< \nu <$ 30 cm $^{-1}$ (Mishima, 1983). $n$ is assumed to be constant 
for $\nu < $ 50 cm $^{-1}$, and $\log{k}$ is linearly interpolated between 30 cm$^{-1}$ and 50 cm $^{-1}$.
The black dashed lines represent $n$ and $k$ with the intra-grain porosity $\phi$ of 0.5.
The blue lines represent $n$ and $k$ for amorphous water ice at 106 K (Curtis et al., 2005).

\item
Fig.~6. Extinction efficiency, $Q_{\rm ext}$, single scattering albedo, $\omega$, and emissivity, $\epsilon_{\rm model}$,
as a function of  wavenumber calculated from the Mie theory and the Hapke's model. 
Different colors represent different grain sizes. In the panels of $\omega$ and $\epsilon_{\rm model}$,
the black lines represent  a size distribution case with $r_{\rm min} = 1$ $\mu$m, $r_{\rm max} = 1$ cm, and $p = 3$.
(a) parameters calculated without diffraction removal ($Q_{\rm ext0}$,  $\omega_0$, and $\epsilon_{\rm model}$ calculated with $\omega_0$), 
(b) parameters calculated with diffraction removal, and (c) parameters calculated with diffraction removal and the intra-grain porosity $\phi$ of 0.5.
For Figs. 6a and 6b, $\phi = 0$.

\item
Fig.~7. Contour of $\tilde{R}_2$ on the $p$ vs. $r_{\max}$ plane.
A cross represents the location of the best-fit values where $\tilde{R}_2 = 0$.

\item
Fig.~8. Emissivities from observations $\epsilon_1$ (blue) and $\epsilon_2$ (black) and modeled emissivity 
$\epsilon_{\rm model}$ (red) with the best fitted $r_{\rm max}$ and $p$ (shown in the parenthesis) for
the inner C ring, the outer B ring, the Cassini division, and the A ring.
For each ring, one high temperature case and one low temperature case are shown.
The fitted temperatures and the fraction of the warm portion are shown in the panels.

\item
Fig.~9. Estimated $p$ and $r_{\rm max}$ for all temperatures and  all rings.  

\item
Fig.~10. Effective radius $r_{\rm eff}$ (Eq.~(\ref{eq:reff})) and median radius $r_{\rm med}$ (Eq.~(\ref{eq:rmed})) calculated from $p$ and $r_{\rm max}$ in Fig.~9.

\end{description}

\clearpage

\begin{figure}

\begin{center}
\includegraphics[width=.75\textwidth]{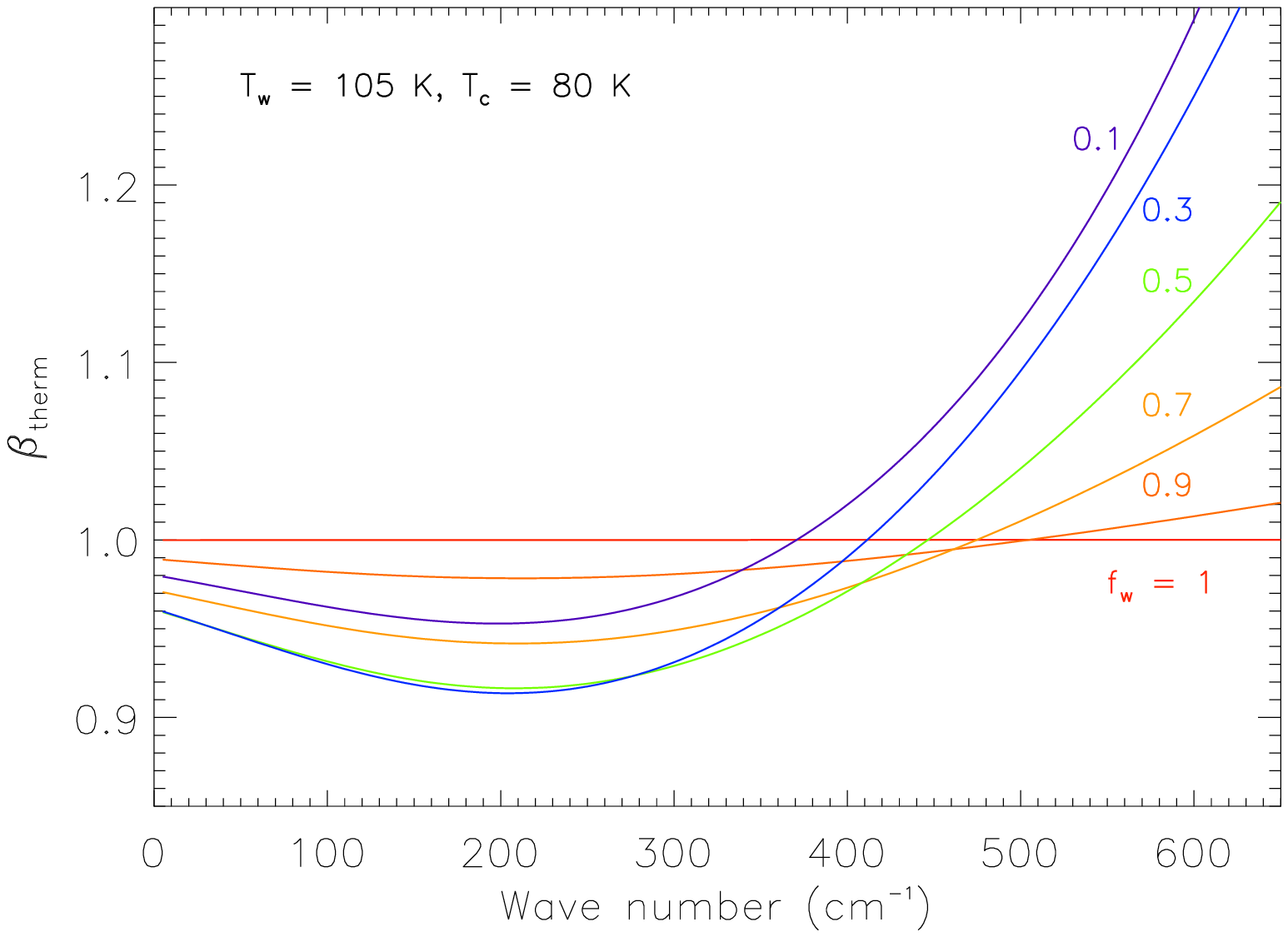}

\end{center}

Fig.~1.

\end{figure}

\clearpage

\begin{figure}

\begin{center}
\includegraphics[width=.8\textwidth]{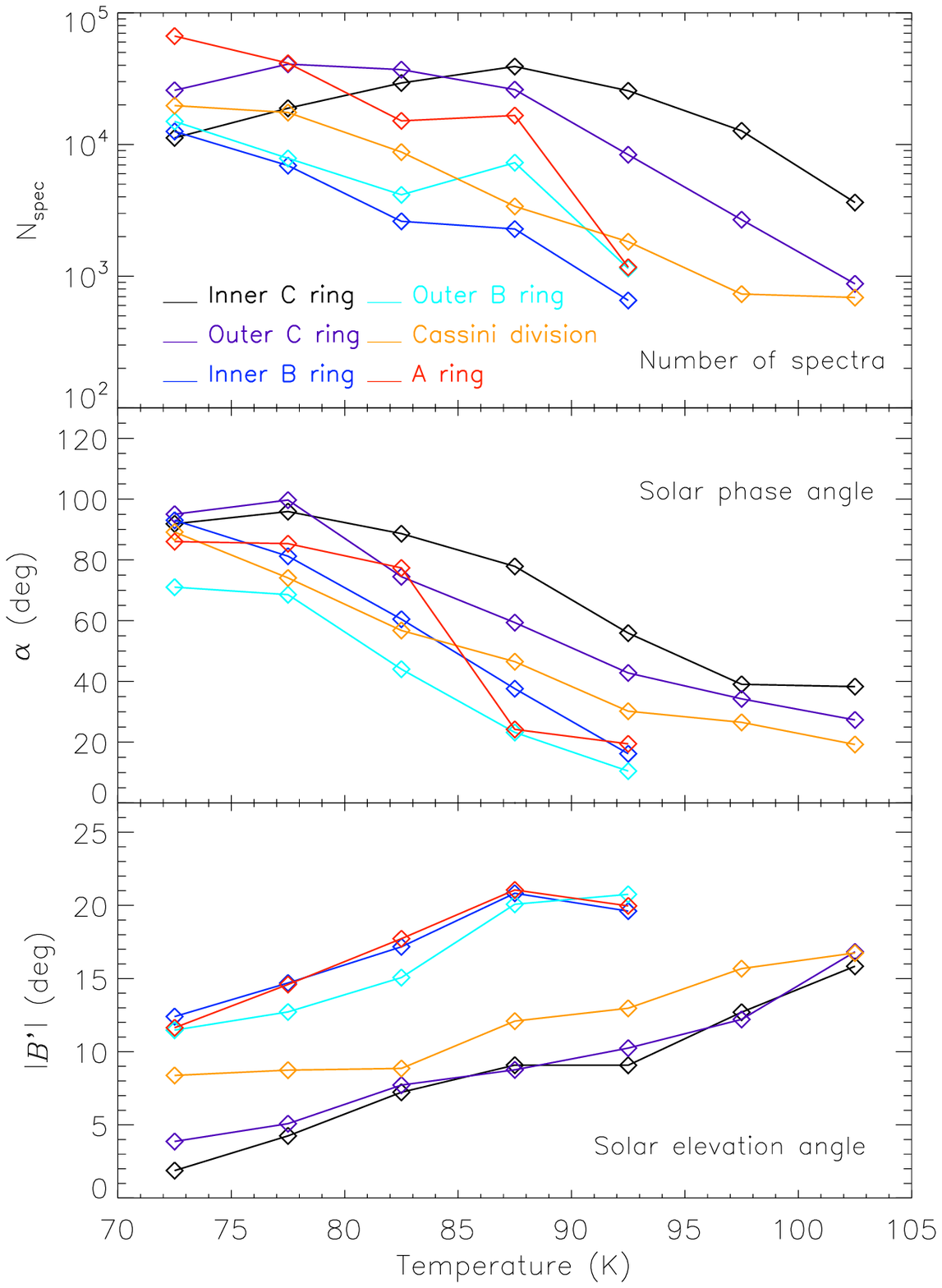}

\end{center}

Fig.~2. 

\end{figure}

\clearpage

\begin{figure}

\begin{center}
\includegraphics[width=.75\textwidth]{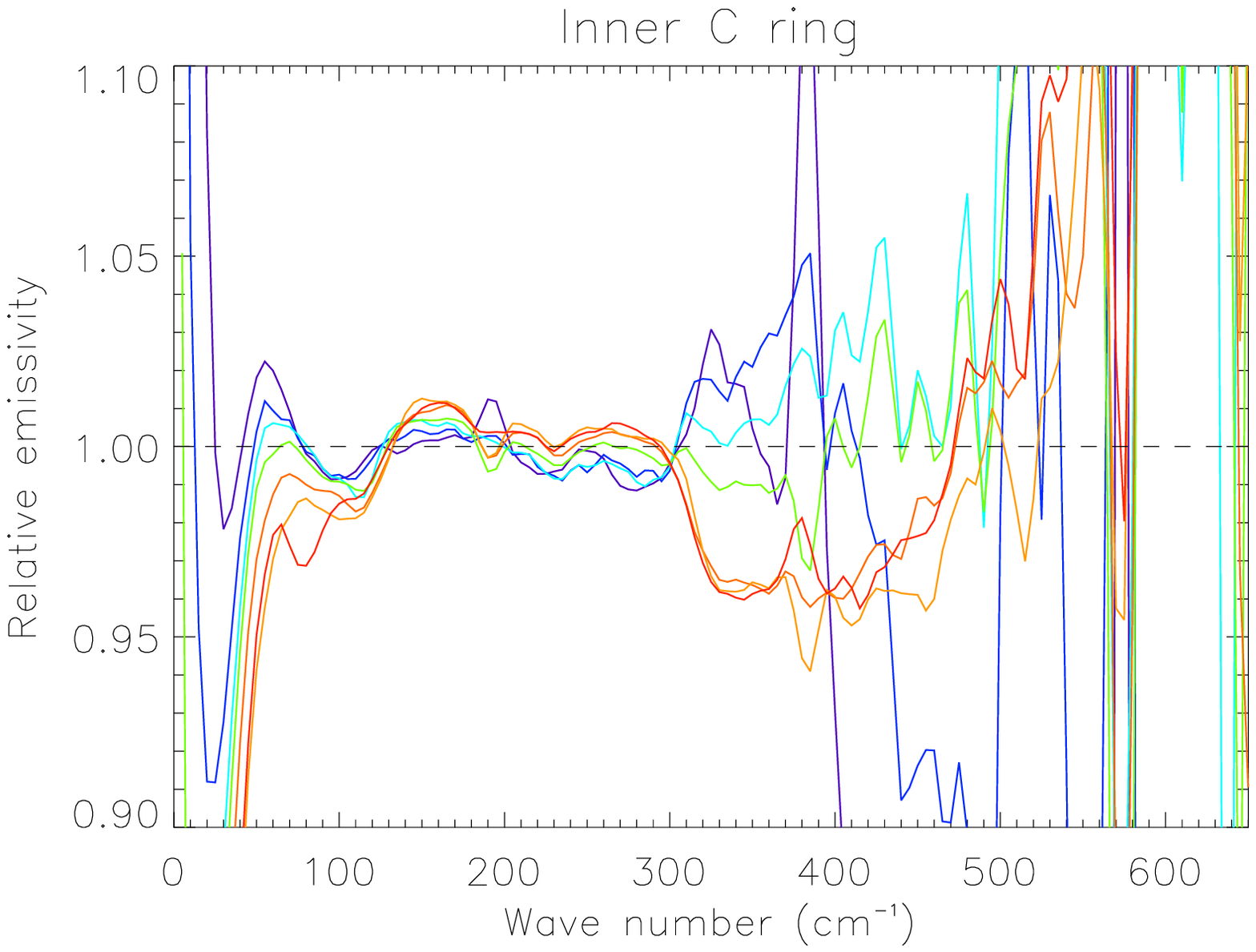}
\includegraphics[width=.75\textwidth]{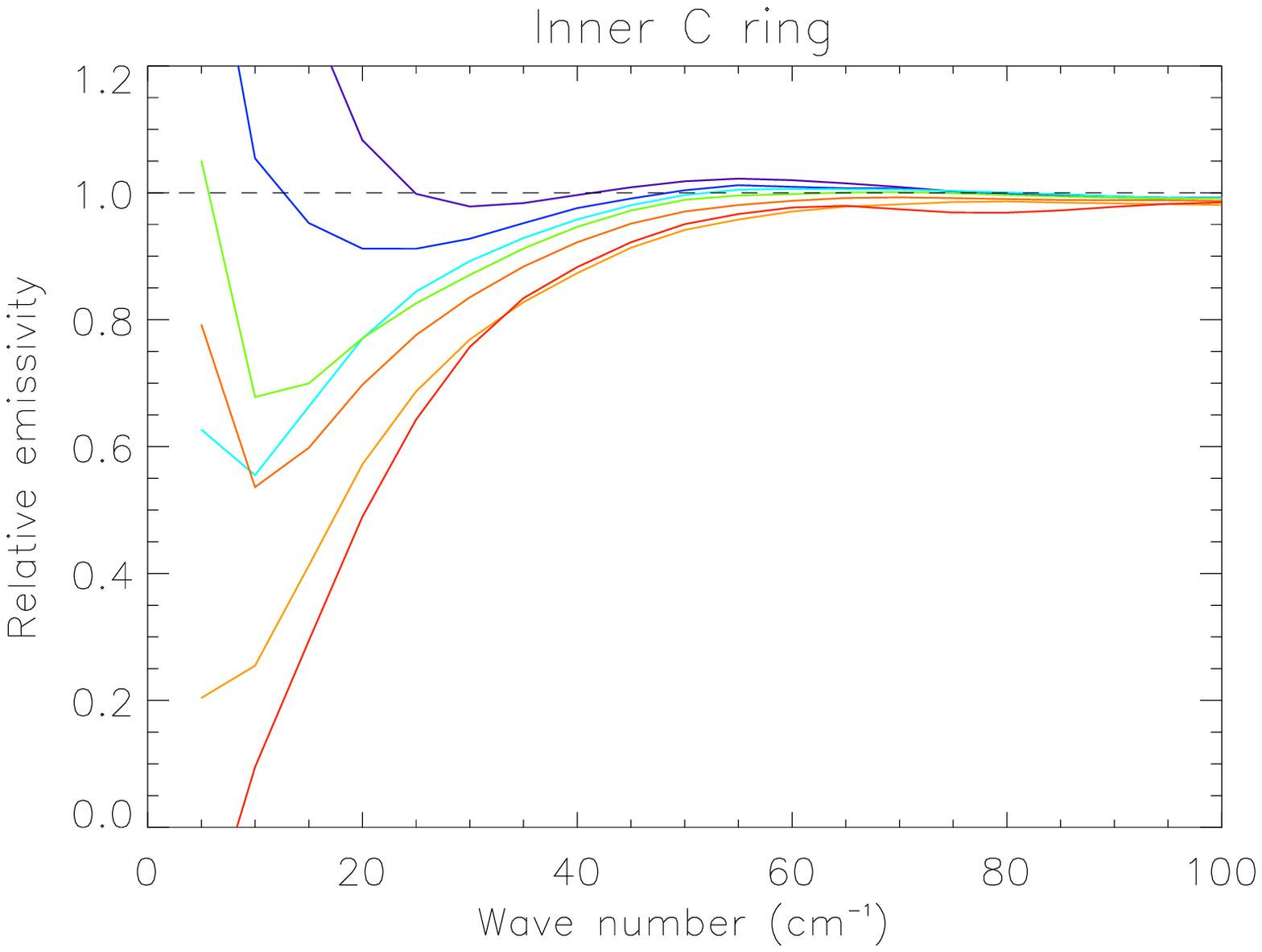}
\end{center}

Fig.~3. 

\end{figure}

\clearpage

\begin{figure}

\begin{center}
\includegraphics[width=.75\textwidth]{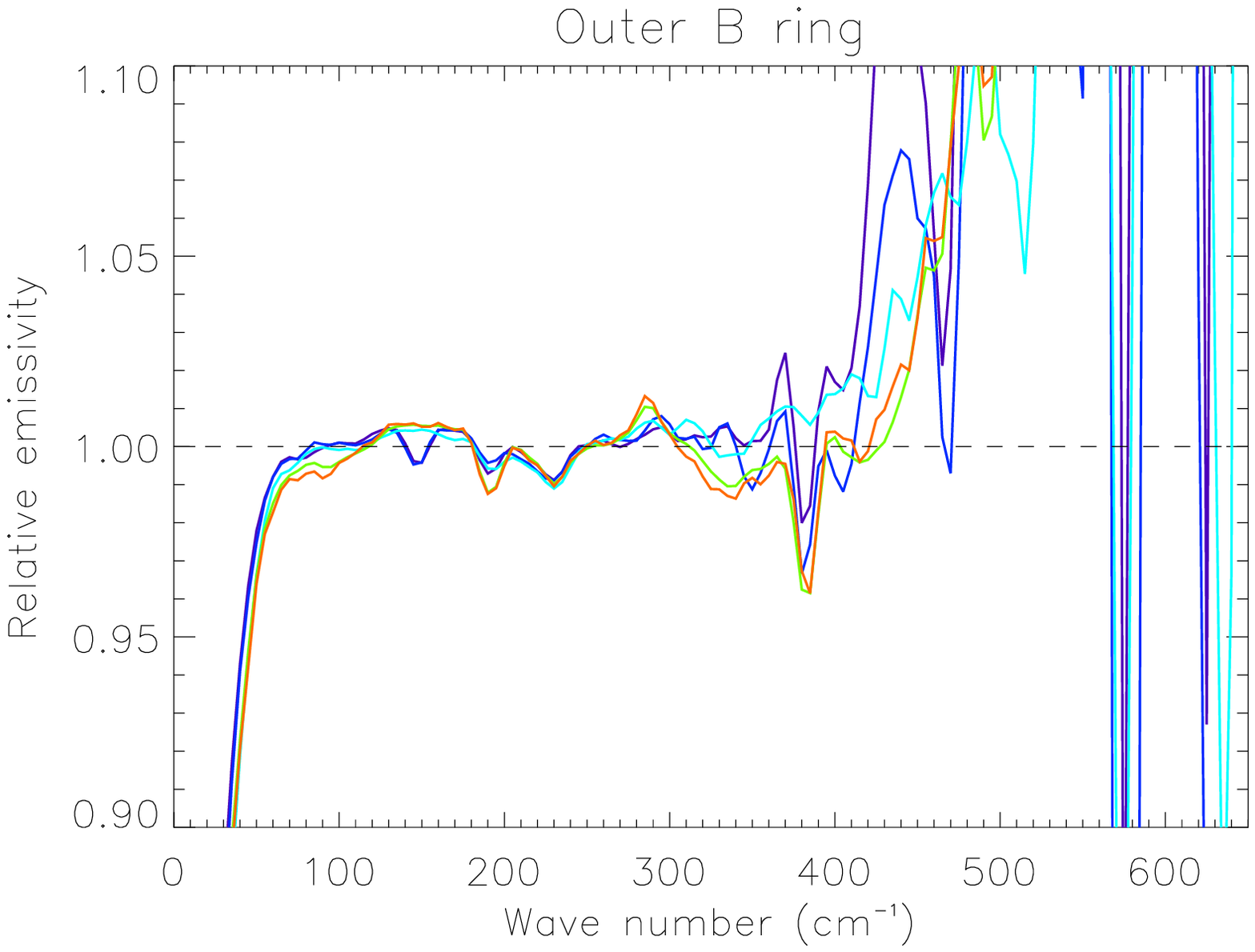}
\includegraphics[width=.75\textwidth]{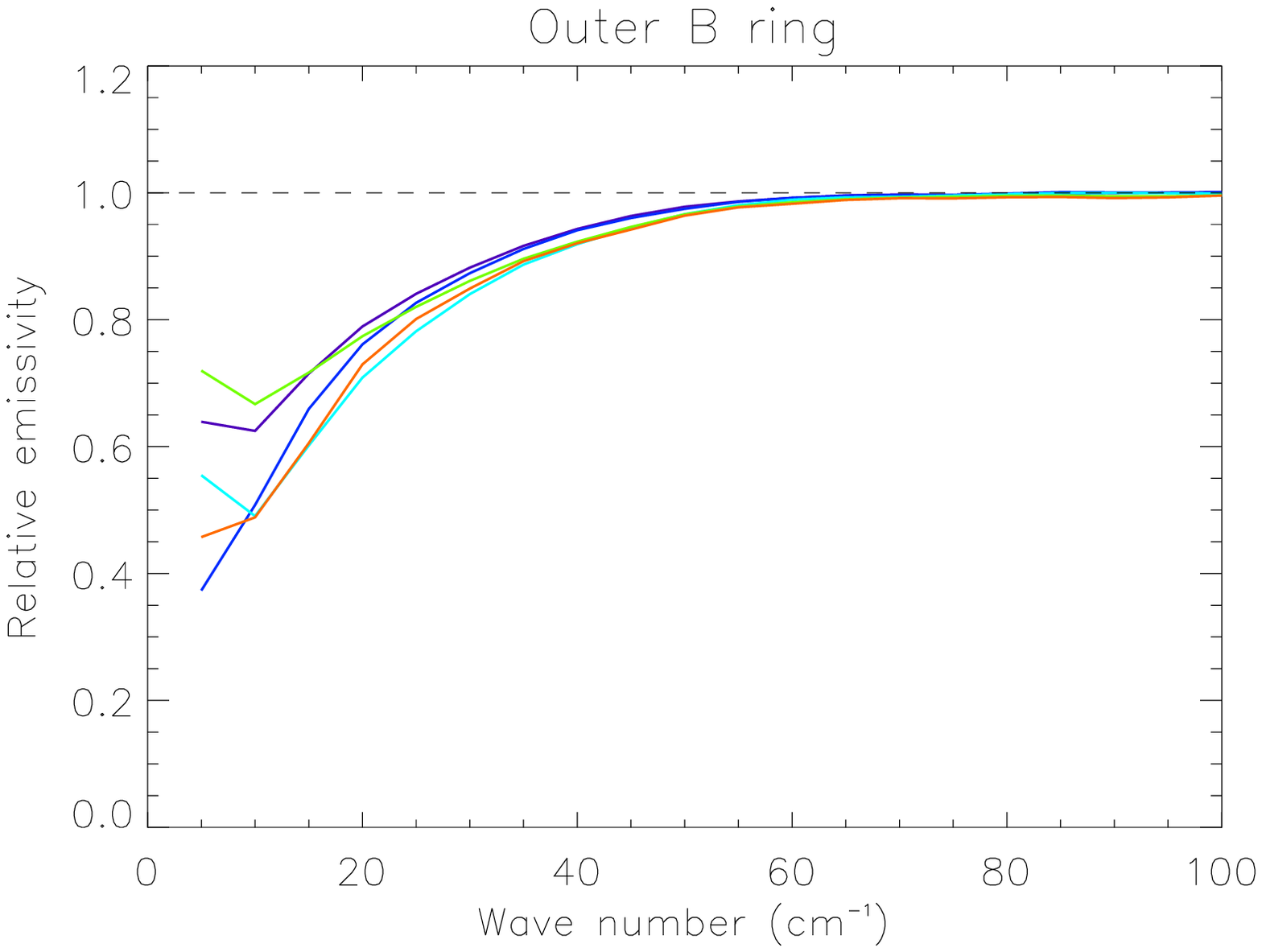}
\end{center}

Fig.~4. 

\end{figure}

\clearpage

\begin{figure}

\begin{center}
\includegraphics[width=.8\textwidth]{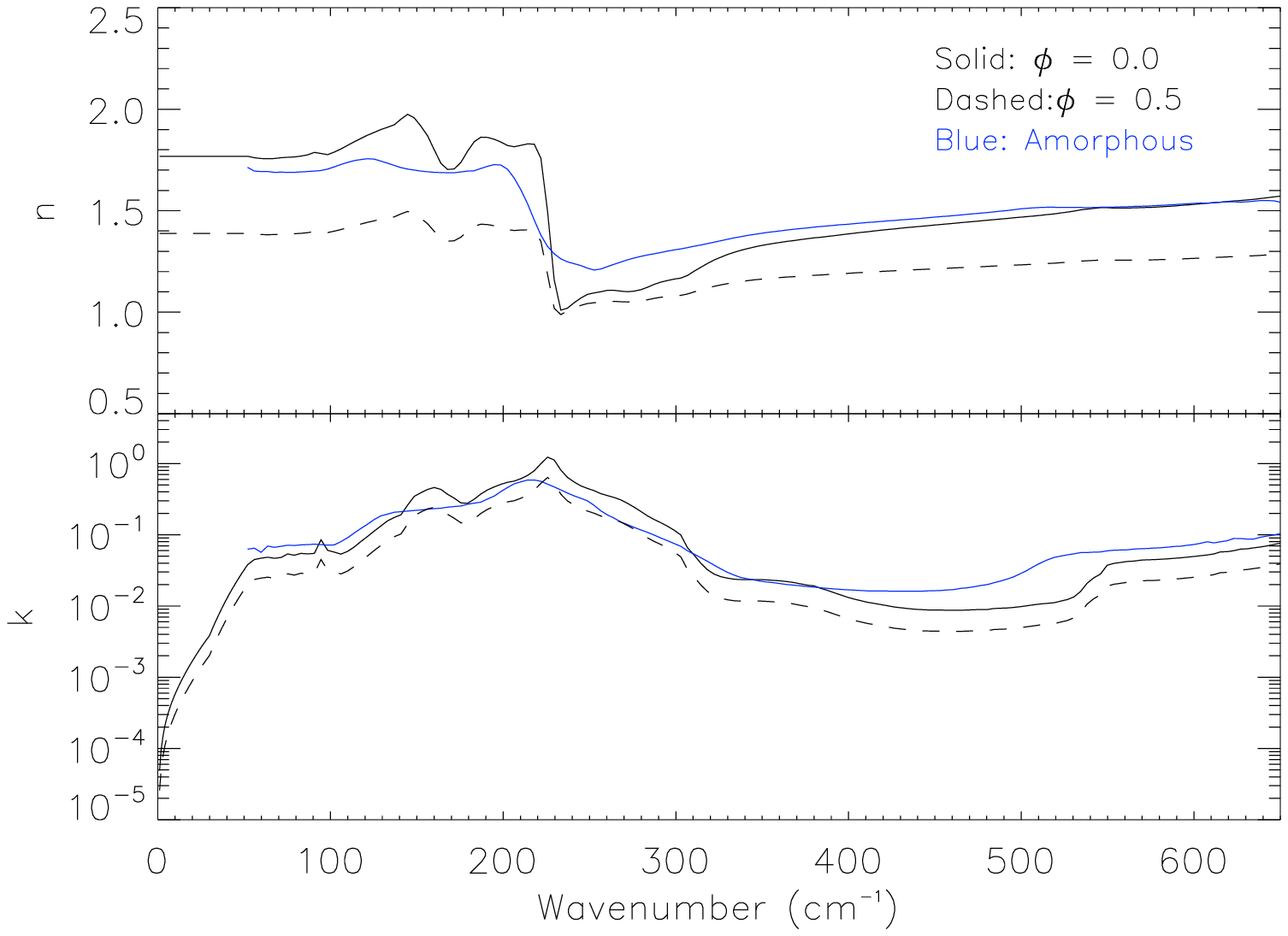}
\end{center}

Fig.~5.

\end{figure}

\clearpage

\begin{figure}

\begin{center}
\includegraphics[width=.7\textwidth]{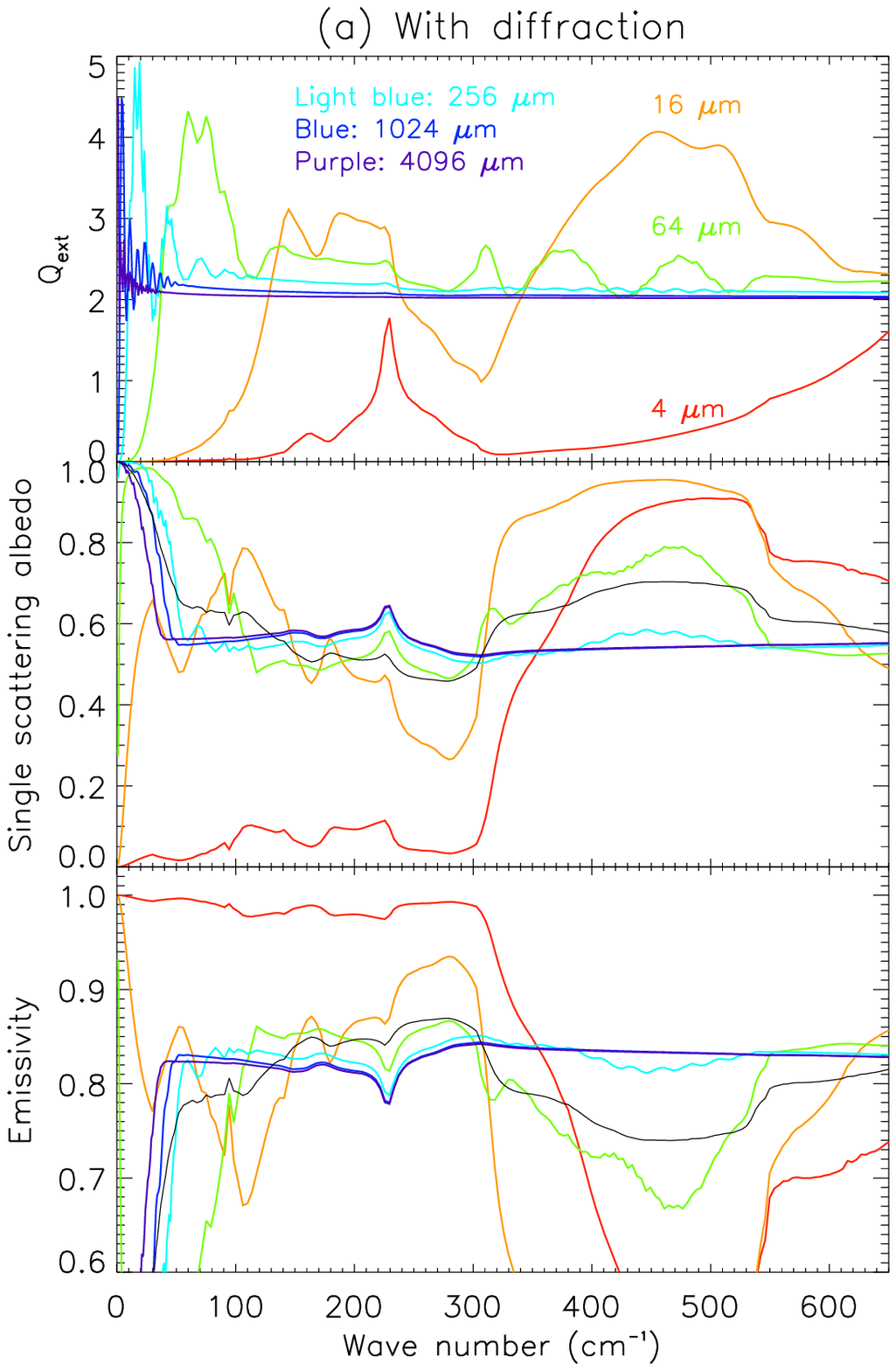}
\end{center}

Fig.~6. 

\end{figure}

\clearpage

\begin{figure}

\begin{center}
\includegraphics[width=.7\textwidth]{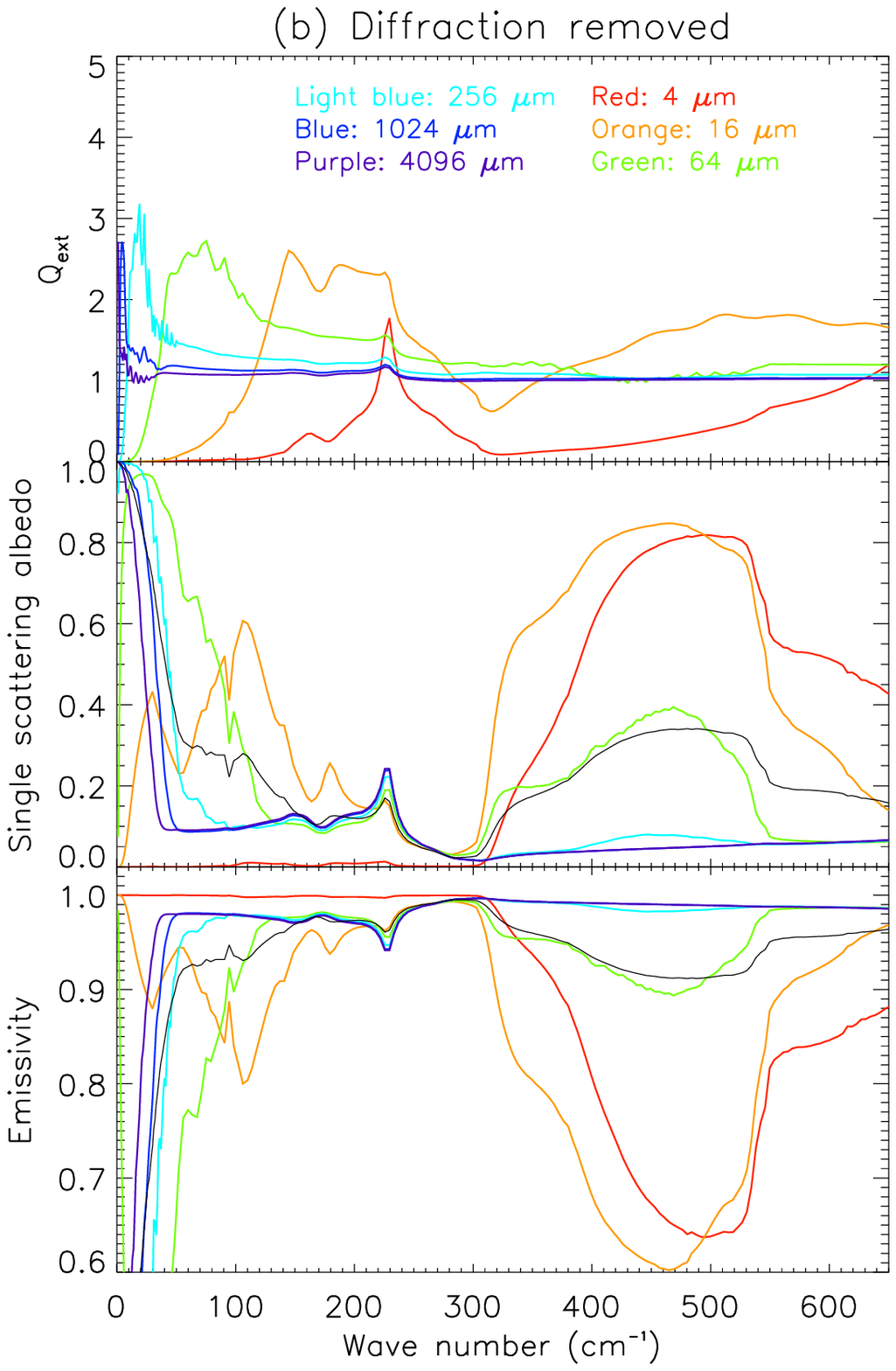}
\end{center}

Fig.~6. -continue 

\end{figure}

\clearpage

\begin{figure}

\begin{center}
\includegraphics[width=.7\textwidth]{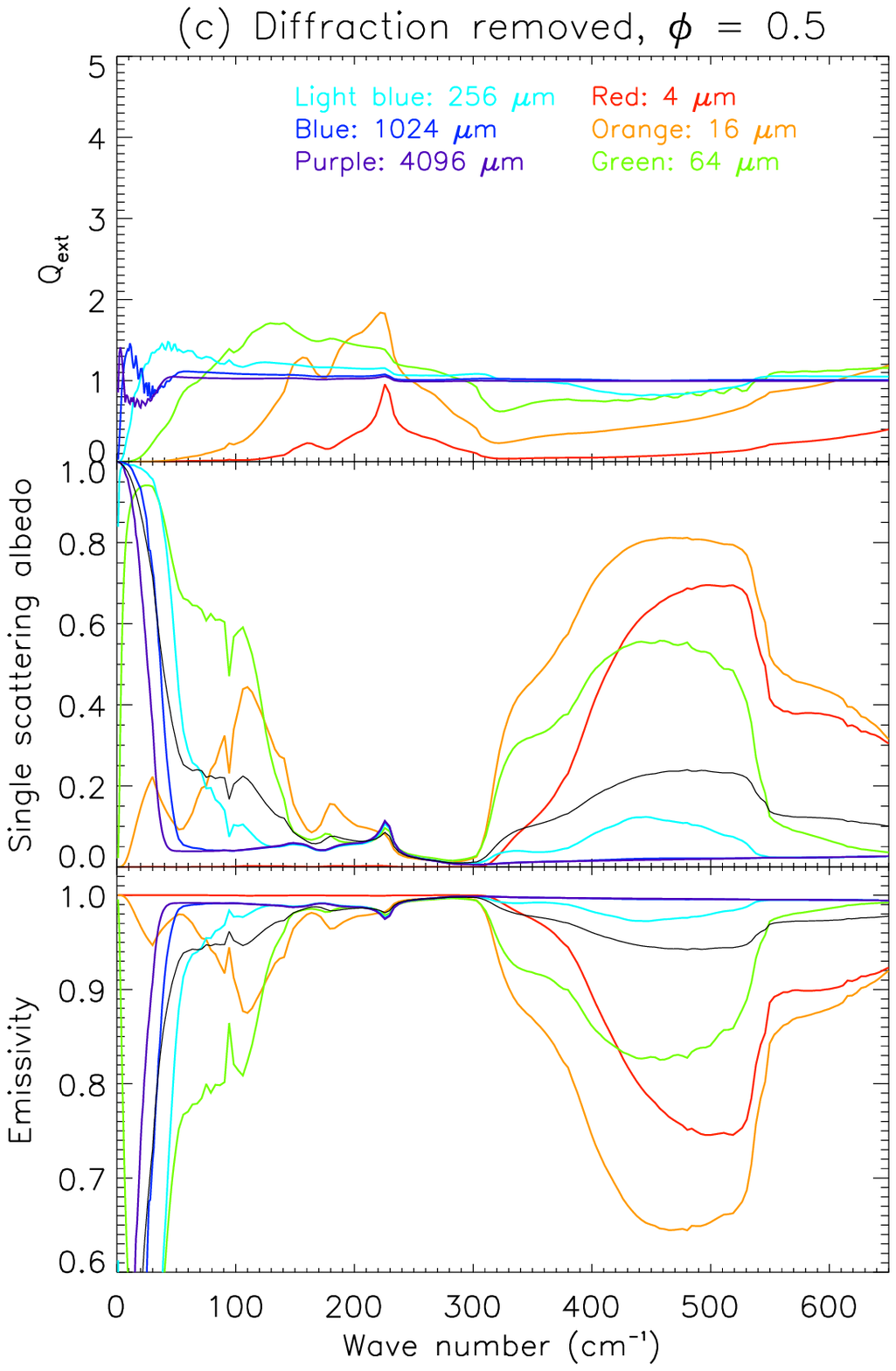}
\end{center}

Fig.~6. -continue 

\end{figure}

\clearpage

\begin{figure}

\begin{center}
\includegraphics[width=.6\textwidth]{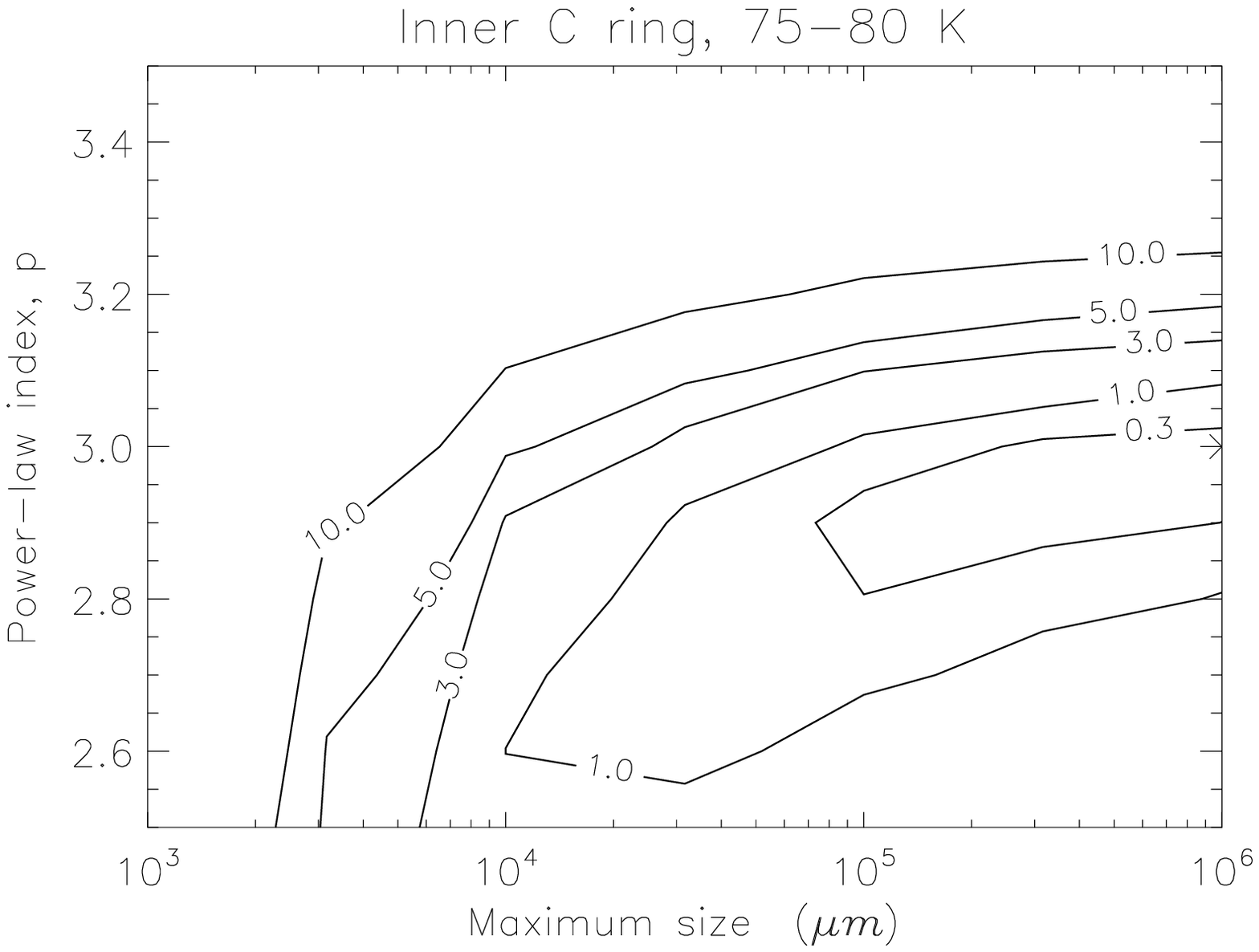}
\includegraphics[width=.6\textwidth]{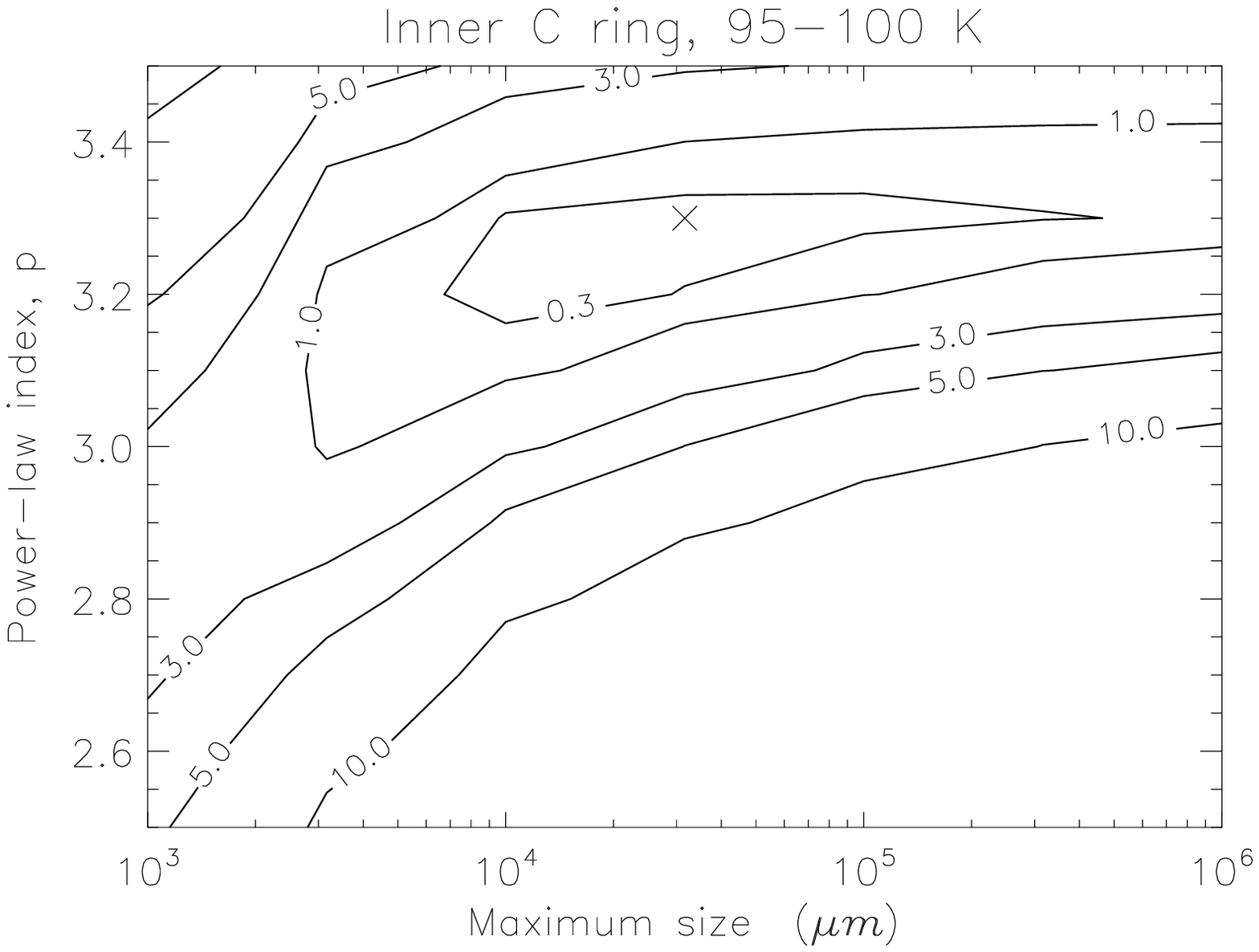}
\includegraphics[width=.6\textwidth]{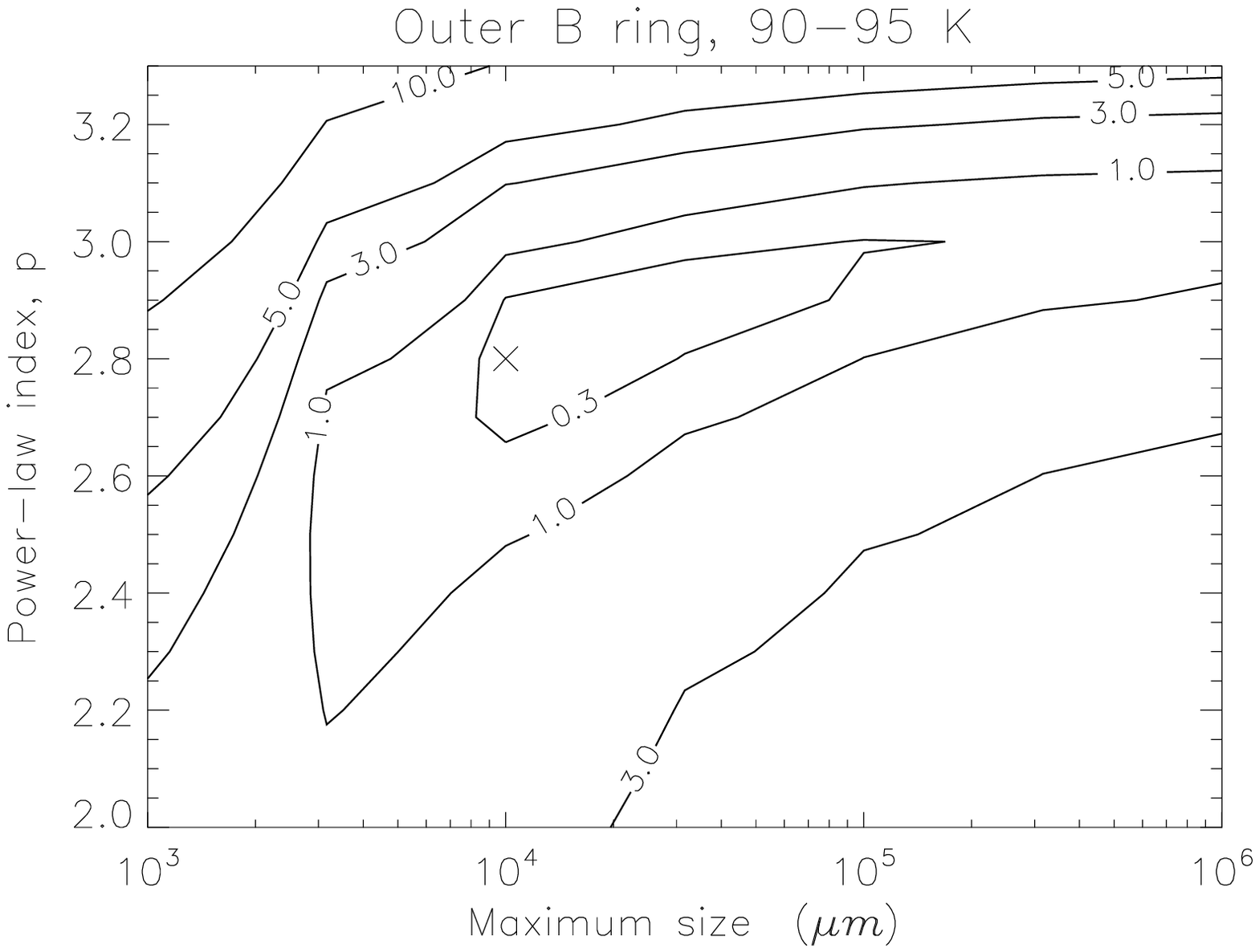}
\end{center}

Fig.~7. 

\end{figure}

\clearpage

\begin{figure}

\begin{center}
\includegraphics[width=.8\textwidth]{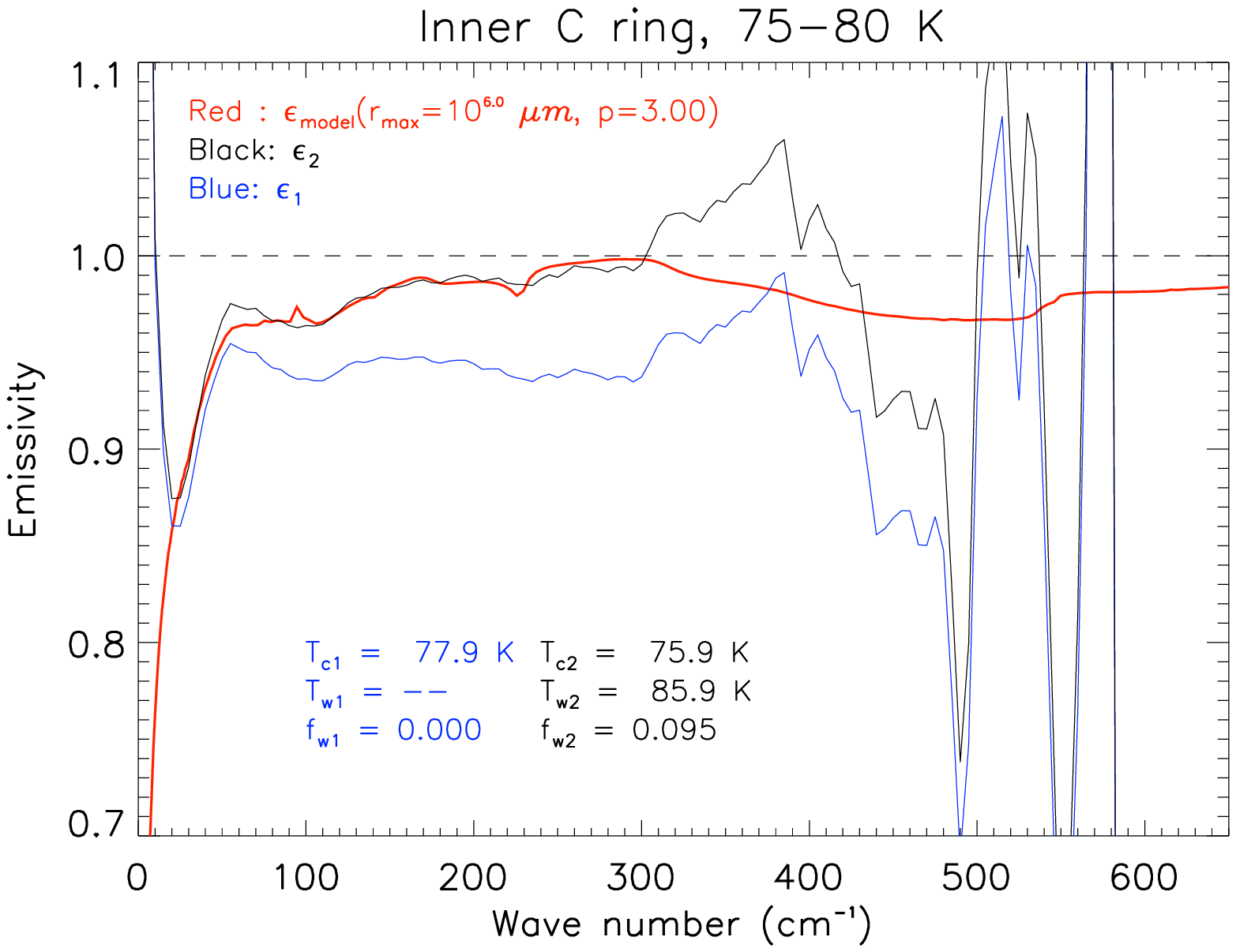}
\includegraphics[width=.8\textwidth]{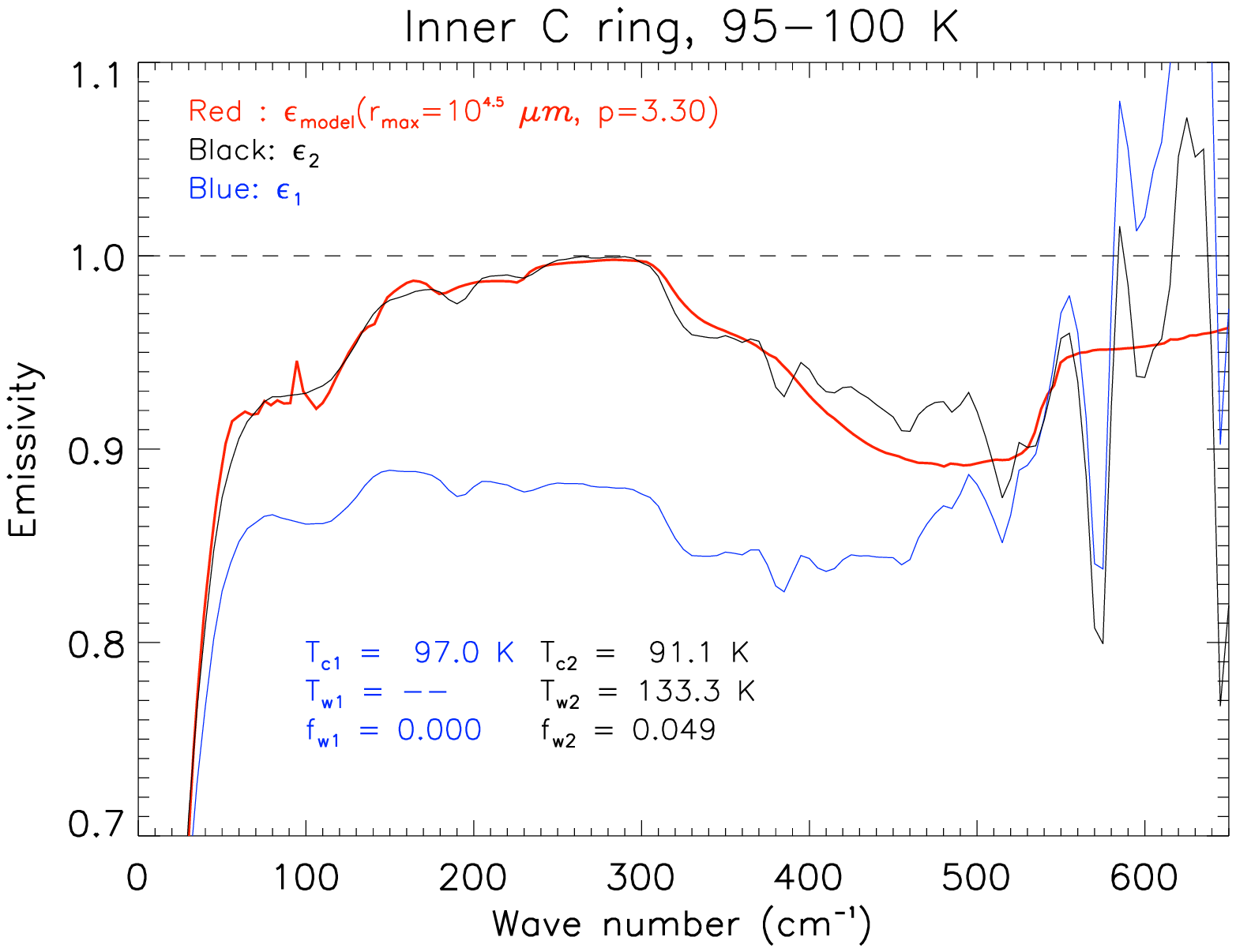}

\end{center}

Fig.~8. 

\end{figure}

\clearpage

\begin{figure}

\begin{center}
\includegraphics[width=.8\textwidth]{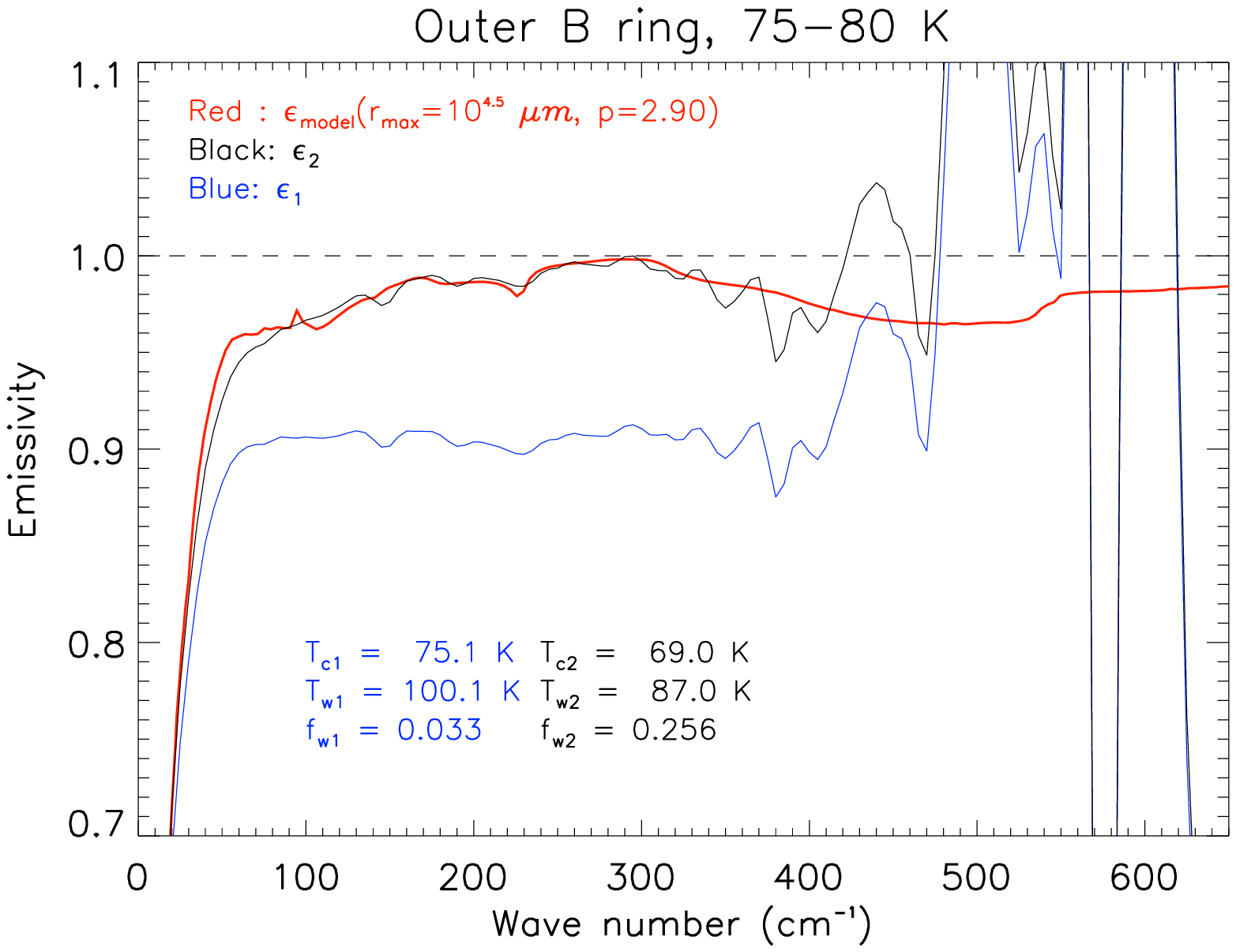}
\includegraphics[width=.8\textwidth]{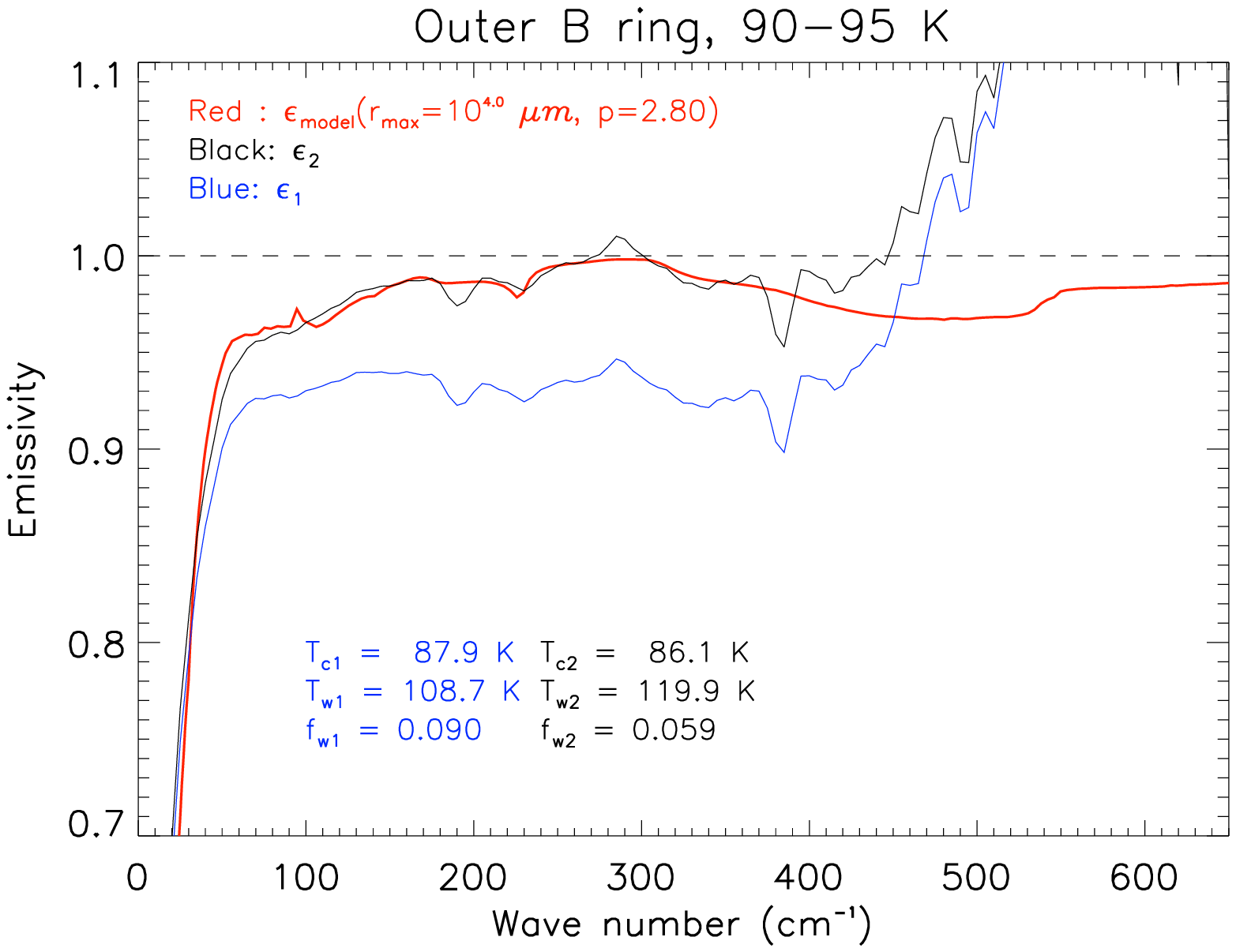}

\end{center}

Fig.~8. -continue

\end{figure}

\clearpage

\begin{figure}

\begin{center}
\includegraphics[width=.8\textwidth]{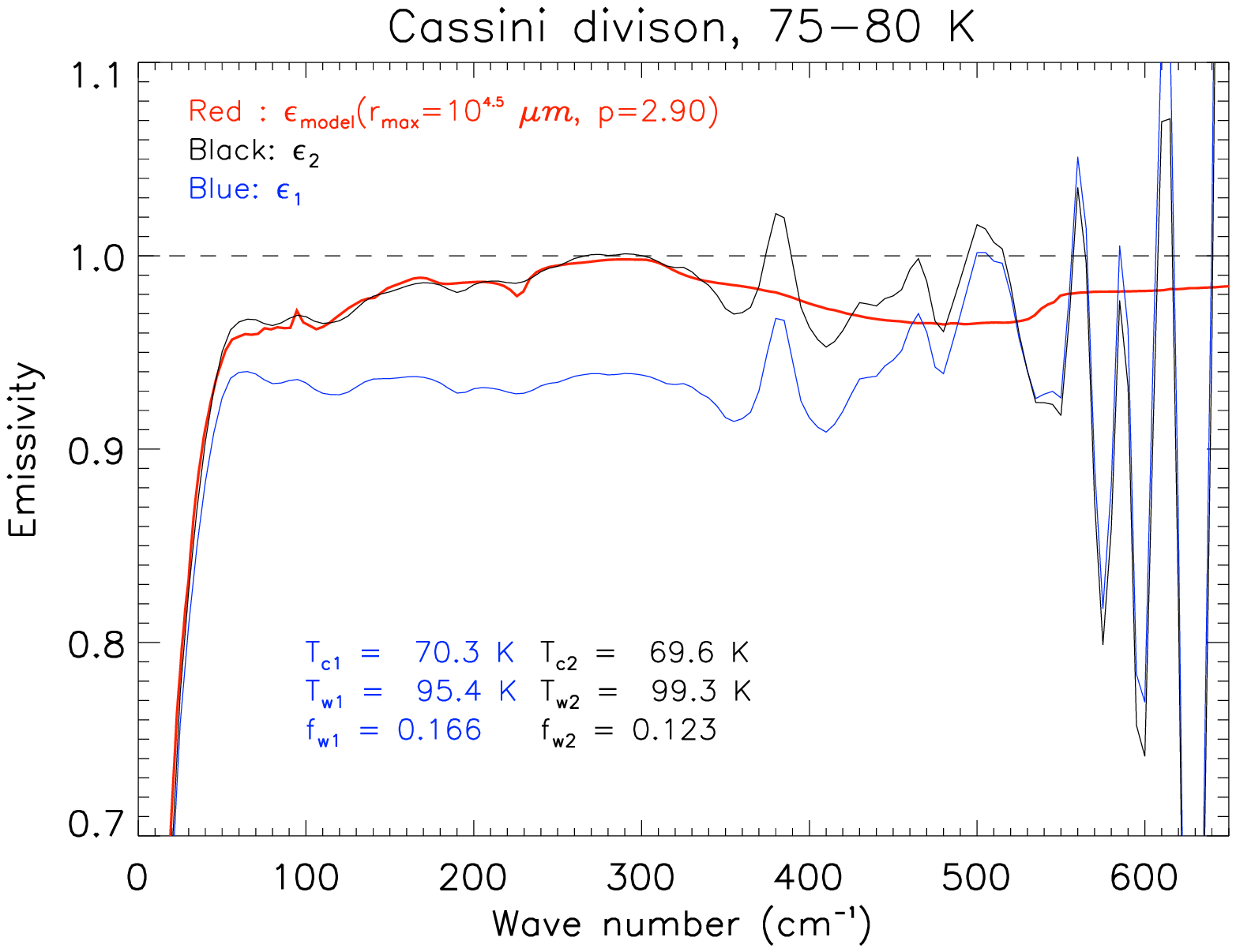}
\includegraphics[width=.8\textwidth]{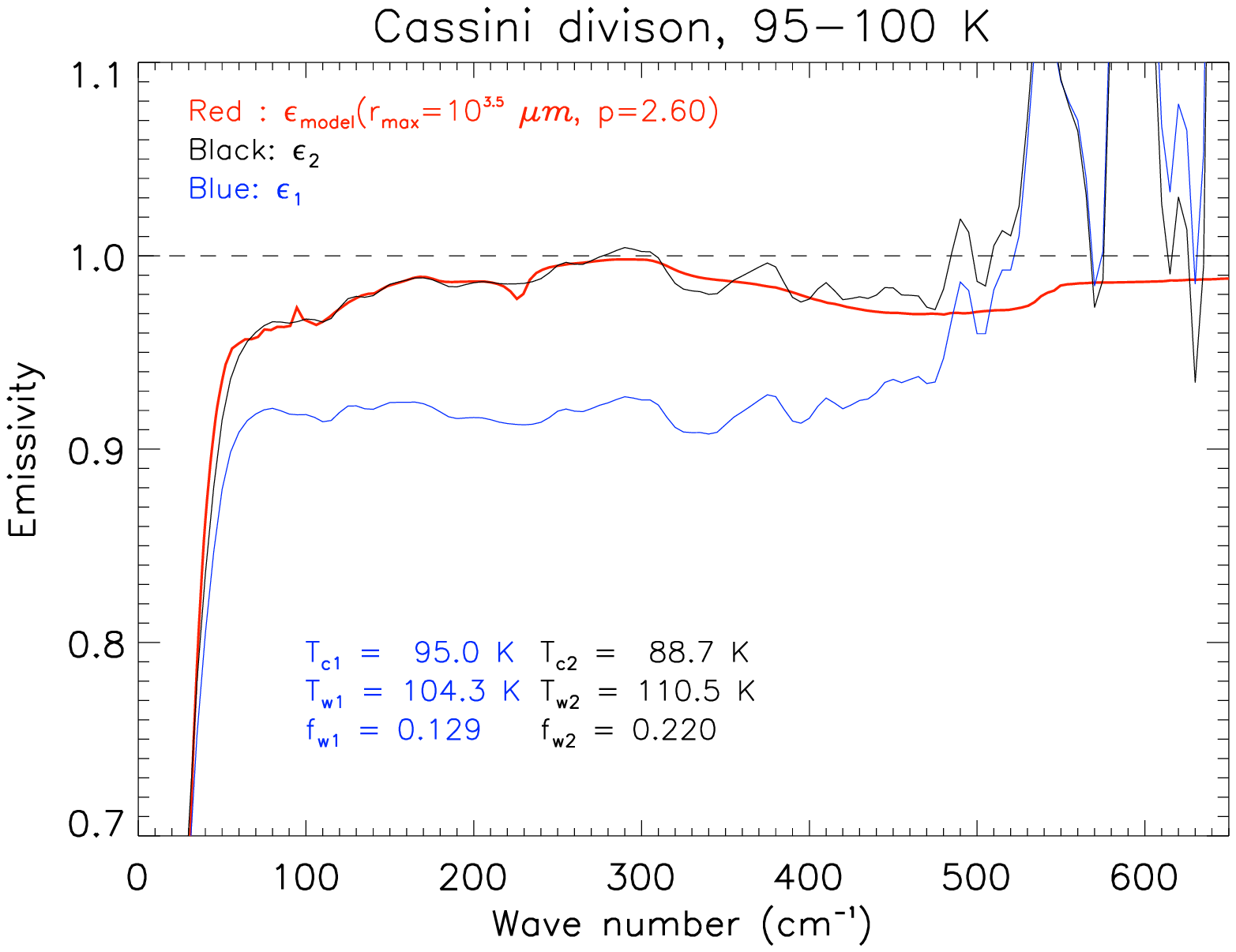}

\end{center}

Fig.~8. -continue

\end{figure}

\clearpage

\begin{figure}

\begin{center}
\includegraphics[width=.8\textwidth]{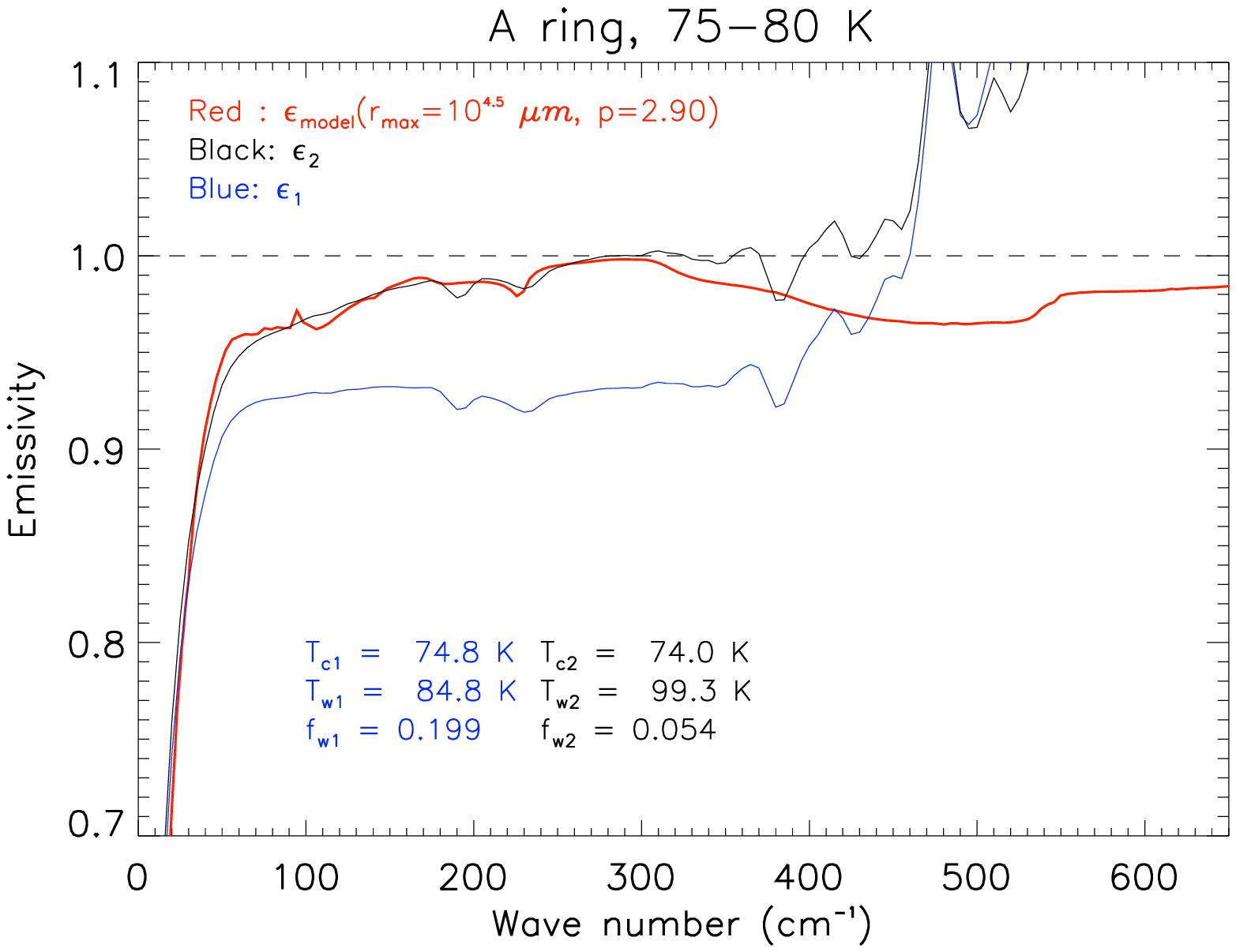}
\includegraphics[width=.8\textwidth]{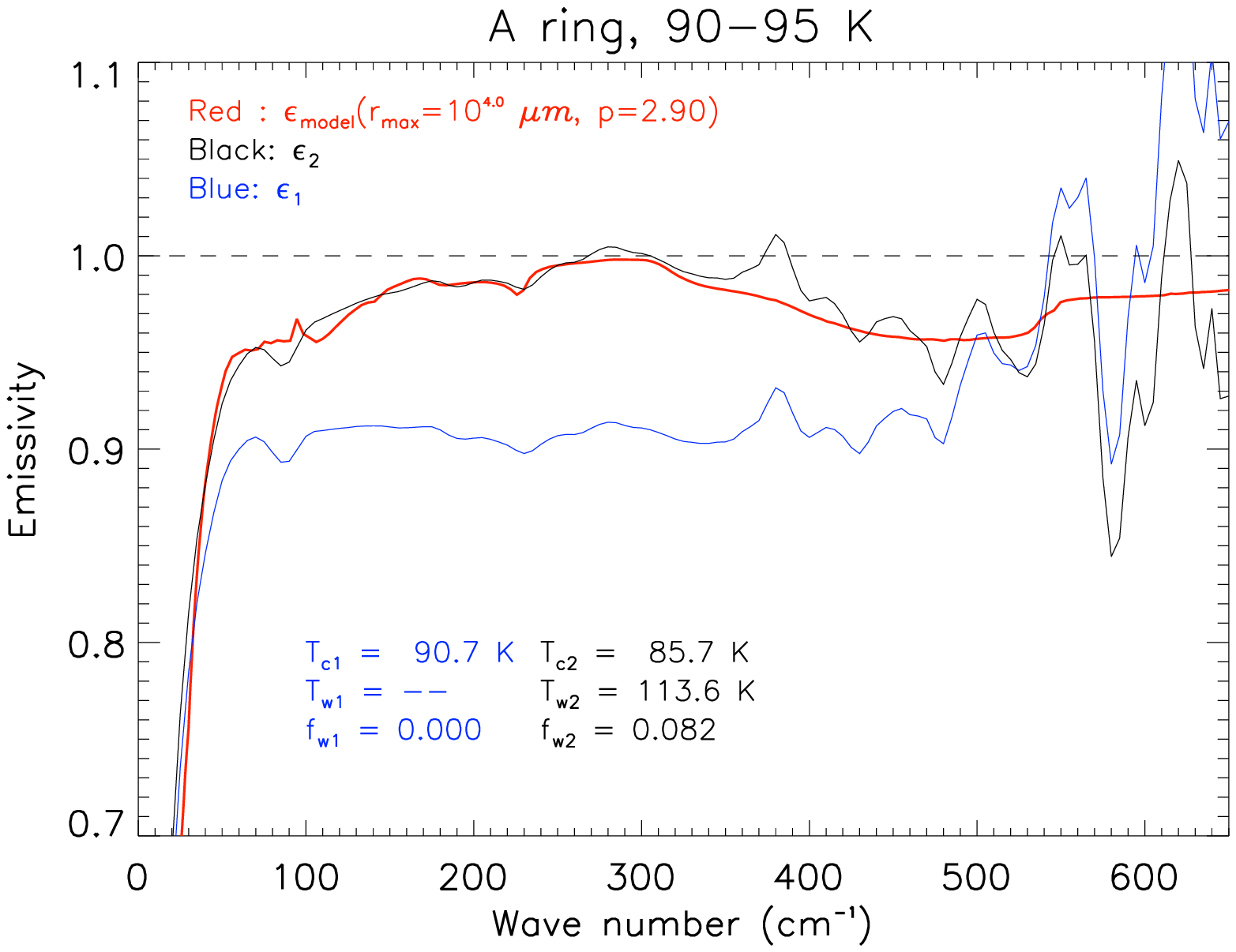}

\end{center}

Fig.~8. -continue

\end{figure}

\clearpage

\begin{figure}

\begin{center}
\includegraphics[width=.8\textwidth]{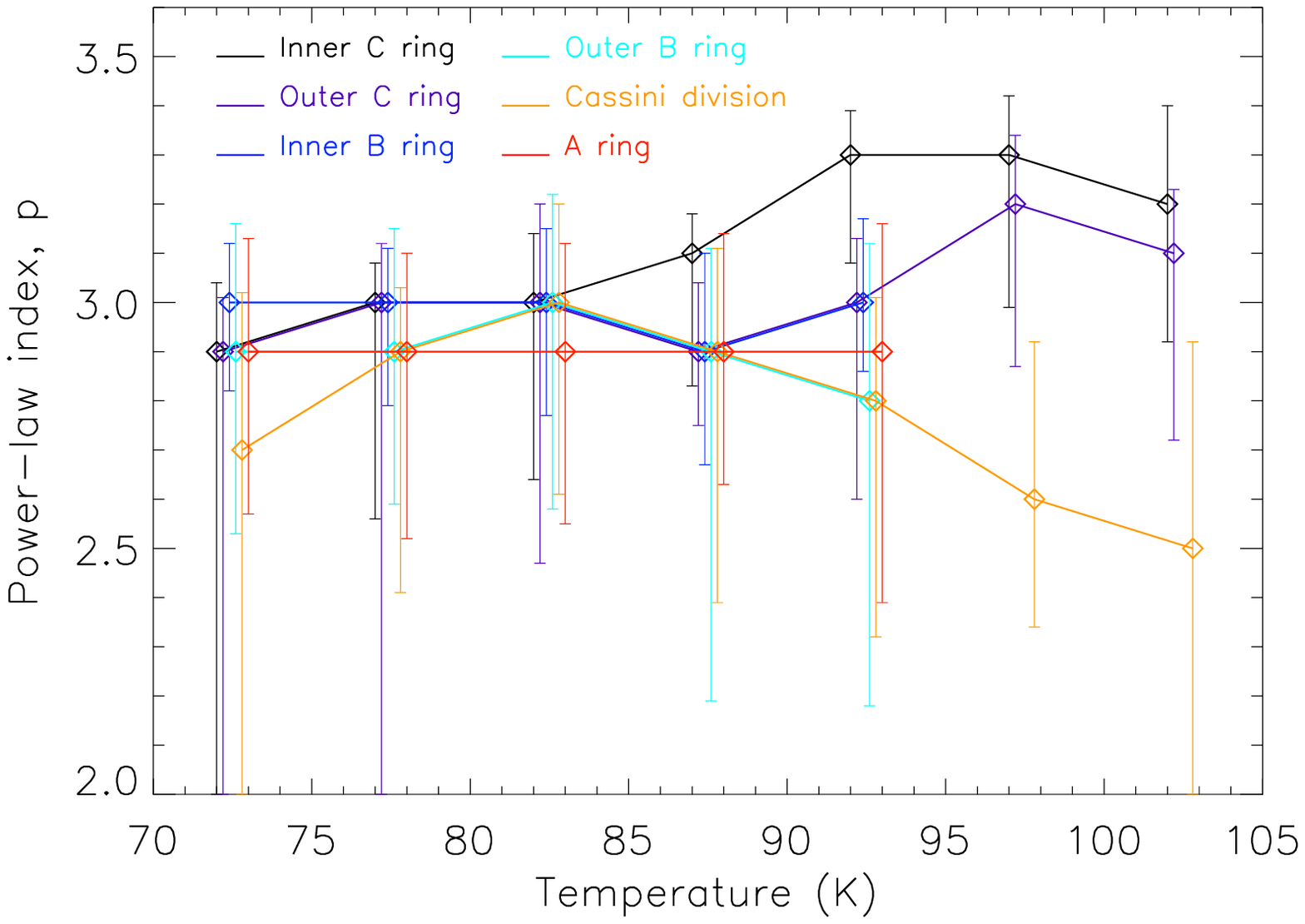}
\includegraphics[width=.8\textwidth]{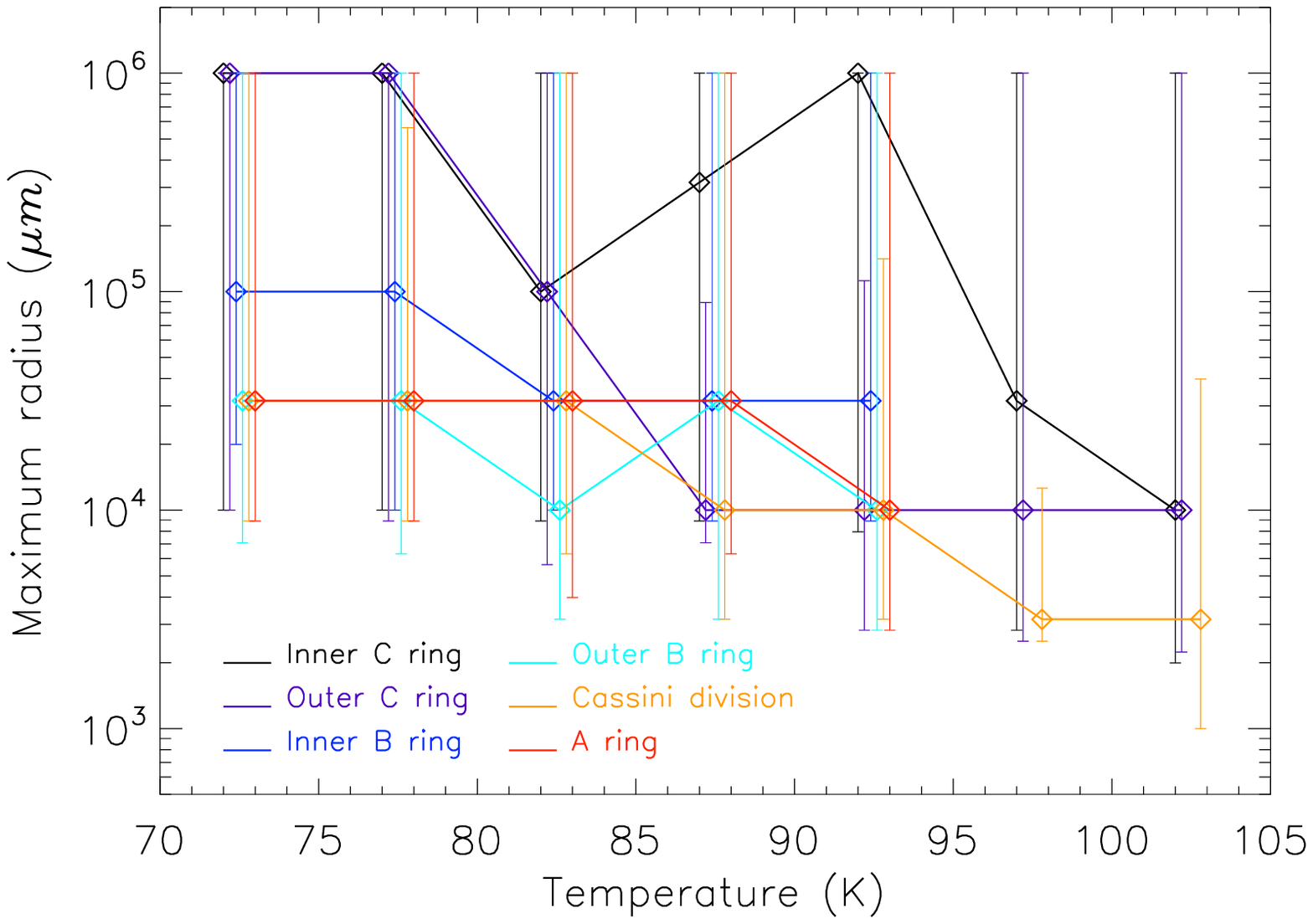}

\end{center}

Fig.~9. 

\end{figure}

\clearpage

\begin{figure}

\begin{center}

\includegraphics[width=.8\textwidth]{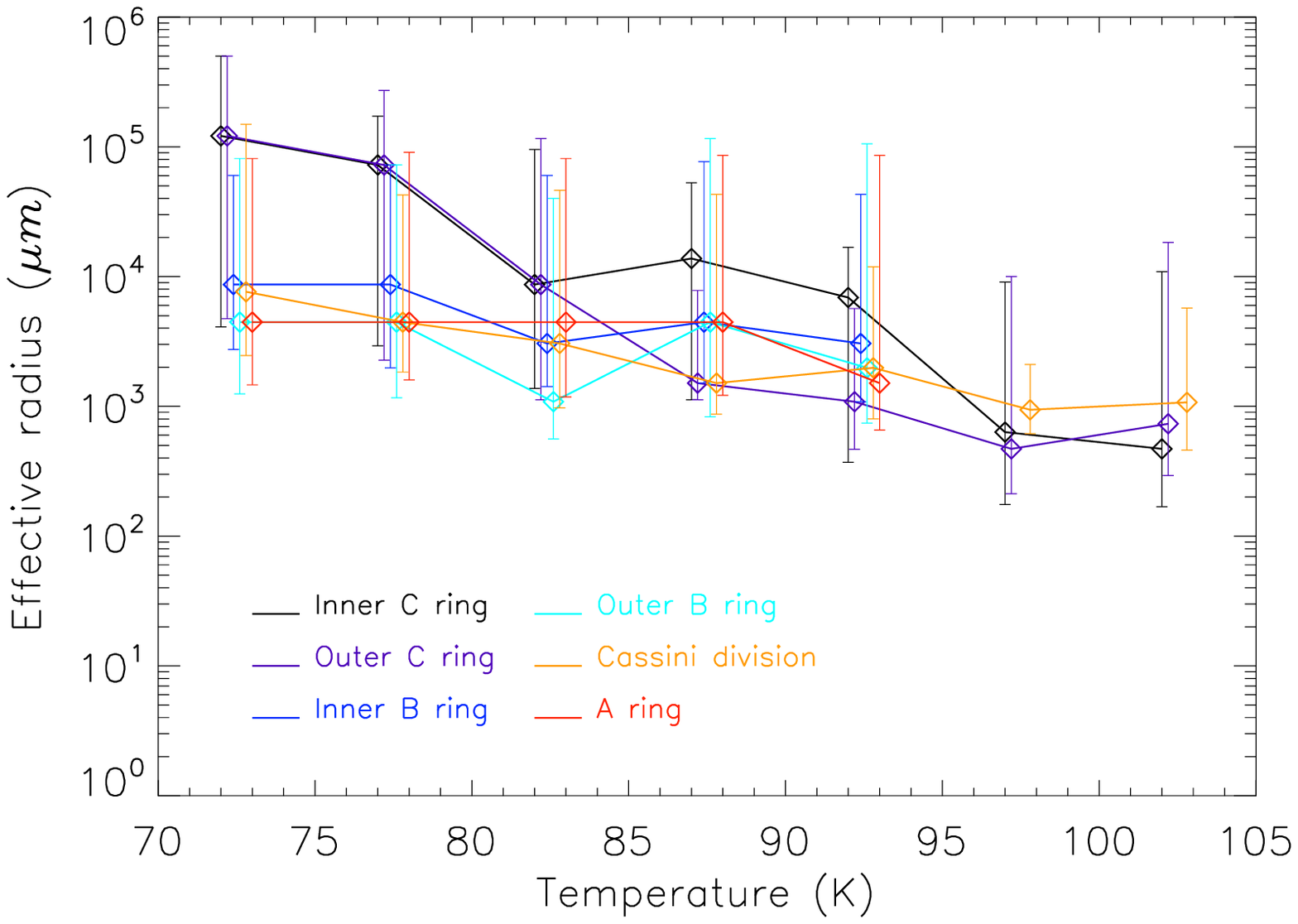}
\includegraphics[width=.8\textwidth]{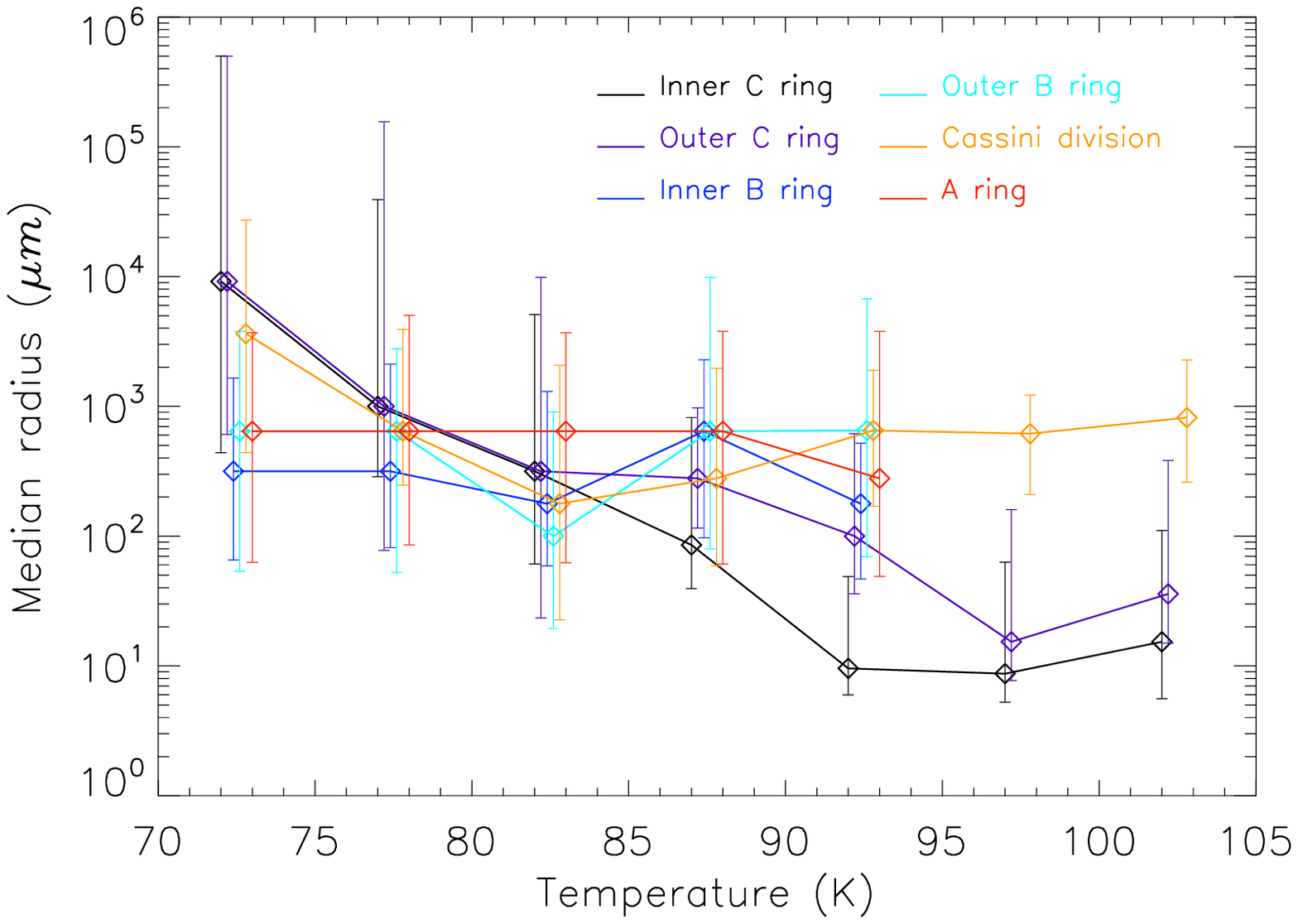}
\end{center}

Fig.~10.

\end{figure}

\clearpage

\end{document}